\documentclass[12pt,aps,prd,preprint,tightenlines,%superscriptaddress,
amsmath,amssymb,showpacs,nofootinbib
]{revtex4}
\usepackage{amsmath}         % \
\usepackage{amssymb}         %  | AMS Packages for math
\usepackage{amsfonts}        % /
\usepackage{graphicx}        % Graphics
\usepackage{wasysym}

\usepackage{lipsum}        % block of text (formatting test)
\usepackage{color}         % \color{...}, colored text
\usepackage{enumerate}		% 
\graphicspath{{figures/}}	% set directory for figures

\renewcommand{\tilde}{\widetilde}   % tilde over characters
\renewcommand{\email}[1]{\href{mailto:#1}{\texttt{#1}}}

\let\oldenumerate\enumerate
\renewcommand{\enumerate}{
  \oldenumerate
  \setlength{\itemsep}{1pt}
  \setlength{\parskip}{0pt}
  \setlength{\parsep}{0pt}
}

\let\olditemize\itemize
\renewcommand{\itemize}{
  \olditemize
  \setlength{\itemsep}{1pt}
  \setlength{\parskip}{0pt}
  \setlength{\parsep}{0pt}
}

\newcommand{\PRE}[1]{{#1}}  % Use if preprint style

\newcommand{\kev}{\text{keV}}
\newcommand{\mev}{\text{MeV}}
\newcommand{\gev}{\text{GeV}}
\newcommand{\tev}{\text{TeV}}

\newcommand{\cm}{\text{cm}}

\renewcommand{\eqref}[1]{Eq.~(\ref{#1})}

\newcommand{\secref}[1]{Sec.~\ref{sec:#1}}

\newcommand{\figref}[1]{Fig.~\ref{fig:#1}}
\newcommand{\figsref}[2]{Figs.~\ref{fig:#1} and \ref{fig:#2}}

\newcommand{\sla}[1]{\not{\! #1}}

\newcommand{\lag}{\mathcal L}

\usepackage[colorlinks=true,citecolor=blue,linkcolor=blue,
urlcolor=blue,hypertexnames=false]{hyperref}

\usepackage{microtype}				% for clean typography

\begin{document}

\preprint{UCI-TR-2016-18}
\preprint{UCR-TR-2017-FLIP-001}

\title{\PRE{\vspace*{1.in}}
{
Dark Photons from Captured Inelastic Dark Matter Annihilation: Charged Particle Signatures
}
\PRE{\vspace*{.15in}}}

\author{Jordan Smolinsky\footnote{\email{jsmolins@uci.edu}}
}
\affiliation{Department of Physics and Astronomy, University of
  California, Irvine, California 92697, USA 
\PRE{\vspace*{.15in}}
}

\author{Philip Tanedo\footnote{\email{flip.tanedo@ucr.edu}}
}
\affiliation{Department of Physics and Astronomy, University of
  California, Riverside, California 92521, USA
\PRE{\vspace*{.15in}}
}

%\date{\today}

\begin{abstract}
\PRE{\vspace*{.2in}}
\noindent
The dark sector may contain a dark photon that kinetically mixes with the Standard Model photon, allowing dark matter to interact weakly with normal matter. In previous work we analyzed the implications of this scenario for dark matter capture by the Sun. Dark matter will gather in the core of the Sun and annihilate to dark photons. These dark photons travel outwards from the center of the Sun and may decay to produce positrons that can be detected by the Alpha Magnetic Spectrometer (AMS-02) on the ISS. We found that the dark photon parameter space accessible to this analysis is largely constrained by strong limits on the spin-independent WIMP-nucleon cross section from direct detection experiments. In this paper we build upon previous work by considering the case where the dark sector contains two species of Dirac fermion that are nearly degenerate in mass and couple inelastically to the dark photon. We find that for small values of the mass splitting $\Delta \sim 100 ~\kev$, the predicted positron signal at AMS-02 remains largely unchanged from the previously considered elastic case while constraints from direct detection are relaxed, leaving a region of parameter space with dark matter mass $100 ~\gev \lesssim m_X \lesssim 10 ~\tev$, dark photon mass $1 ~\mev \lesssim m_{A'} \lesssim 100 ~\mev$, and kinetic mixing parameter $10^{-9} \lesssim \varepsilon \lesssim 10^{-8}$ that is untouched by supernova observations and fixed target experiments but where an inelastic dark sector may still be discovered using existing AMS-02 data.

\end{abstract} 

%95.35.+d Dark matter
%14.70.Pw Other gauge bosons
%95.55.Vj Neutrino, muon, pion, and other elementary particle detectors;
%  cosmic ray detectors

\maketitle

\section{Introduction}
One of the most important open questions in physics is the nature of dark matter. Despite a large and growing international experimental program its particle properties are still unknown, inspiring physicists to produce models of dark matter that naturally avoid detection by existing analyses, to invent new detector technologies, and to discover new ways to repurpose existing detectors. In recent years so-called ``secluded dark matter'' models \cite{Pospelov:2007mp}, wherein dark matter interacts with the Standard Model through a light mediator particle, have gained attention. These models are appealing in part because the correct abundance can be obtained by annihilation to mediators in the early universe while the mediator coupling to the Standard Model can be made parametrically small to evade experimental exclusions. Direct detection experiments still place stringent constraints on the simplest secluded dark matter models, and it is natural to ask if variations on these models can avoid direct detection bounds while remaining accessible to other experiments.

Inelastic models were initially invoked for just this purpose: to reconcile the experimental tension between the DAMA modulation signal and null results from CDMS \cite{TuckerSmith:2001hy,Chang:2008gd}, in part because even a small mass splitting $\Delta \sim 100 ~\kev$ can weaken upper limits on the spin-independent cross section from nuclear recoils by many orders of magnitude. These models are also appealing for their potential to resolve small scale structure problems if the mass splitting is $\Delta \gtrsim 50 ~\kev$ \cite{Blennow:2016gde}. 

In this manuscript we focus on one such model: a simplified model of a self interacting inelastic dark sector that interacts with the Standard Model by kinetic mixing \cite{Holdom:1985ag,Holdom:1986eq} between a hypothesized massive $U(1)_X$ gauge boson, hereafter referred to as the ``dark photon'' \cite{Kobzarev:1966qya,Okun:1982xi}, and the Standard Model photon. 

In previous work \cite{Feng:2015hja,Feng:2016ijc} we investigated a novel indirect detection signal of the elastic limit of this model: dark matter is captured by astrophysical objects and annihilates into dark photons. These dark photons then stream outwards and decay to Standard Model particles that may be detected by existing experiments. We found that dark matter capture and annihilation within the Earth could produce unique signals in the IceCube Neutrino Observatory over a large, currently unconstrained region of parameter space with dark matter masses $100 ~\gev \lesssim m_X \lesssim 10 ~\tev$. We discovered further that the same process occurring in the Sun, which we term ``dark sunshine,'' may result in energetic positrons detectable by the Alpha Magnetic Spectrometer (AMS-02), providing a probe of the dark sector unconstrained by dark photon exclusions from fixed target experiments and supernova observations, with $100 ~\gev \lesssim m_X \lesssim 10 ~\tev$, $1 ~\mev \lesssim m_{A'} \lesssim 100 ~\mev$, and $10^{-10} \lesssim \varepsilon \lesssim 10^{-8}$. This region of parameter space exhibits a large overlap with the parameter space excluded by large noble liquid direct detection experiments. We now ask whether inelastic models that obviate these exclusions could still be found through dark sunshine.

This is not an obvious question. Inelastic models avoid direct detection constraints due to the kinematics of endothermic scattering. Dark matter that falls into the Earth's gravitational well will typically not be energetic enough to scatter. We may suspect, on the other hand, that the stronger gravitational influence of the Sun will allow these interactions to occur, even at mass splittings that prohibit scattering in terrestrial liquid xenon experiments. In such a case we expect that the solar dark matter capture and annihilation rates are only mildly suppressed by the addition of a mass splitting, rendering a dark sunshine search as potentially the only way to observe dark matter in our solar system.

To answer this question, we perform a complete analysis of dark matter capture and annihilation in the Sun with this model, including kinematic suppression of dark matter capture relative to the elastic case, modification of the annihilation rate due to Sommerfeld enhancement in inelastic models, relaxation of competing limits from direct detection, and the effect of the Sun's magnetic field on the positron signal at AMS-02. We find that even with mass splittings large enough to weaken constraints from LUX by two orders of magnitude, the capture and annihilation processes are only weakly affected. As a result, there exist regions of parameter space untouched by other probes of the dark sector, and favored by small scale structure observations, where AMS-02 is expected to see tens or hundreds of energetic positrons produced by boosted dark photon decays in existing datasets. For the case of dark matter capture by the Earth, we find that in the case when the inelastic mass splitting is small our results are unchanged from earlier analysis, but that the Earth does not capture dark matter efficiently when the mass splitting is large enough to substantially modify direct detection limits on the model.

Dark matter capture and annihilation in gravitating bodies has been examined before \cite{Freese:1985qw,Press:1985ug,Silk:1985ax,Krauss:1985aaa,Griest:1986yu,Gaisser:1986ha,Gould:1987ir,Gould:1987ju,Gould:1991hx,Gould:1999je} with early works considering the production of neutrinos by captured dark matter annihilation in the Sun, while more recent studies have focused on the production of new particles beyond the Standard Model, of which dark photons are one example \cite{Batell:2009zp,Schuster:2009au,Schuster:2009fc,Meade:2009mu}. The ANTARES neutrino telescope has recently placed limits on a similar scenario to what we have considered, searching for muons and neutrinos coming from mediator decays in a secluded dark matter framework \cite{Adrian-Martinez:2016ujo}. Kouvaris et al. further examine ``darkonium'' bound states in the Sun and argue that AMS-02 may be sensitive to dark photons emitted during the formation of these bound states \cite{Kouvaris:2016ltf}. Inelastic dark matter models have also been previously applied to other astrophysical anomalies \cite{ArkaniHamed:2008qn}, collider searches \cite{Bai:2011jg,Izaguirre:2015zva}, and solar dark matter capture \cite{Menon:2009qj}.

\section{Model}
\label{sec:model}
The hidden sector we consider in this manuscript consists of two Dirac fermions of nearly identical masses interacting inelastically through the gauge boson of a broken $U(1)_X$ gauge symmetry, called the ``dark photon,'' that kinetically mixes with the Standard Model hypercharge boson. Inelastic couplings may be introduced in the fermion sector by diagonalizing a mass matrix that induces maximal mixing between the two species. A model of an inelastic dark sector consisting of Majorana fermions was given in Ref.~\cite{TuckerSmith:2001hy}, and most SUSY-inspired dark sector models share this trait. To comport with our earlier work on dark matter capture and annihilation we have chosen to use a simplified model of inelastic Dirac dark matter. This class of model was first invoked in Ref.~\cite{Harnik:2008uu}. Here we provide an explicit construction to establish notation. Suppose that the dark sector consists of two Dirac fermions, denoted $\psi$ and $\chi$, which carry opposite-sign $U(1)_X$ charges. The relevant dark sector interactions are
\begin{equation}
\lag_\text{dark} \supset g_X \bar{\psi} \sla{A}' \psi - g_X \bar{\chi} \sla{A}' \chi - \left( \bar \psi ~~ \bar \chi \right) \left( \begin{array}{cc}
M & m \\ 
m & M
\end{array}  \right) \left(\begin{array}{c}
\psi \\ 
\chi
\end{array}  \right) \ .
\end{equation}
We assume $M \gg m$. We can now rewrite the Lagrangian in terms of the mass eigenstates $X_{1,2} = \left(\psi \mp \chi\right)/\sqrt{2}$:
\begin{equation}
\lag_\text{dark} \supset g_X \bar{X}_2 \sla{A}' X_1 + g_X \bar{X}_1 \sla{A}' X_2 - (M-m)\bar{X}_1 X_1 - (M+m)\bar{X}_2 X_2 \ .
\end{equation}
If the mass matrix is more general both elastic and inelastic couplings may occur, with arbitrary relative strength. We have chosen to consider the simpler case. We emphasize that these inelastic couplings may be accomplished without the introduction of any new fields beyond a second fermion in the dark sector: the mass matrix consists of terms that are either $U(1)_X$ invariant or that may be produced when the $U(1)_X$ is broken by a dark-charged Higgs field. We assume that the Higgs field responsible for the dark photon mass is heavy enough that its dynamics do not affect the phenomena we will examine. 

The diagonalization of the dark photon--SM gauge Lagrangian has been detailed elsewhere \cite{Feldman:2007wj,Izaguirre:2015eya,Feng:2016ijc}, and results in an effective theory where the dark photon has mixing parameter $\varepsilon$ suppressed couplings to the SM electromagnetic current, and coupling to the weak neutral current further suppressed by $m_{A'}^2/m_Z^2$. Because we consider dark photons of mass below $1 ~\gev$, the latter coupling can be safely neglected, and the effective Lagrangian of our simplified model is
\begin{align}
\begin{split}
\mathcal L &=
-\frac 14 F_{\mu\nu}F^{\mu\nu}
-\frac 14 F'_{\mu\nu}F'^{\mu\nu}
+ \frac{1}{2} m^2_{A'} A'^2
+ \sum_f \bar{f} \left[i \sla{\partial} - q_f e (\sla{A} + \varepsilon \sla{A}') - m_f\right] f
\\
&~ ~~+i \bar{X}_1 \sla{\partial} X_1 + i \bar{X}_2 \sla{\partial} X_2 + g_X \bar X_2\sla{A}' X_1 + g_X \bar X_1 \sla{A}' X_2 - m_X \bar X_1 X_1 \\
&~ ~~- (m_X + \Delta) \bar X_2 X_2 \ .
\end{split}
\label{eq:Lagrangian}
\end{align}
In the above expression we sum over SM fermions $f$ with electric charge $q_f$, $g_X$ is the hidden $U(1)$ gauge coupling, $m_X \equiv M-m$ is the mass of the lighter of the two hidden sector particles, the dominant component of the dark matter, and $\Delta\equiv 2m$ is the mass splitting between the two species. Henceforth in this manuscript, unless there is ambiguity, we will omit subscripts and refer to the lighter species as simply $X$.

This model has five free parameters: $m_X$, $\Delta$, $m_{A'}$, $g_X$, and $\varepsilon$. We assume that $m_X > m_{A'}$. We fix $g_X$ by assuming that the dark matter abundance is set by thermal freeze-out of the process $\bar X X \rightarrow A' A'$, which is the dominant annihilation channel in the case where $\varepsilon \ll g_X$. In the limit where the mass splitting is much smaller than the dark matter mass, which we assume throughout, the thermal relic density is unaffected at the 10\% level by the presence of the more massive state \cite{Griest:1990kh}. 
In order to satisfy the observed abundance as a thermal relic we take \cite{Steigman:2012nb}
\begin{equation}
\alpha_X = 0.035 \left(\frac{m_X}{\tev}\right) \ .
\end{equation}

The dark photon decay rate is given by
\begin{equation}
\Gamma = \frac{1}{\text{Br}(A' \rightarrow e^+ e^-)} \frac{\varepsilon^2 \alpha (m_{A'}^2 + 2 m_e^2)}{3 m_{A'}}\sqrt{1-\frac{4 m_e^2}{m_{A'}^2}} \ .
\end{equation}
Our assumption that $m_X > m_{A'}$ implies that there are no hidden sector states into which the dark photon can decay. The branching ratio to electrons is then $1$ when $2 m_e < m_{A'} < 2 m_\mu$, and at higher masses it is determined from hadron production at colliders \cite{Buschmann:2015awa}.
From this we write the boosted dark photon decay length, which determines whether dark photons produced by dark matter annihilation are able to escape the Sun before decaying and thus produce a positron signal at AMS-02:
\begin{equation}
L = \frac{v \gamma}{\Gamma} =  R_\odot ~\text{Br}(A' \rightarrow e^+ e^-) \left(\frac{1.1 \times 10^{-9}}{\varepsilon}\right)^2 \left(\frac{m_X/m_{A'}}{1000}\right)\left(\frac{100 ~\mev}{m_{A'}}\right) \ ,
\label{eq:decaylength}
\end{equation}
where $R_\odot = 7 \times 10^{10} ~\text{cm} = 4.6 \times 10^{-3} ~\text{au}$ is the radius of the Sun and we have taken the limit $m_e \ll m_{A'} \ll m_X$. 

\section{Experimental Constraints}
\label{sec:exp. constraints}
The possibility of a hidden $U(1)_X$ boson that kinetically mixes with the photon has been extensively investigated, both as part of an interacting dark sector and as a stand-alone extension of the Standard Model. Here we review those experimental constraints relevant for the region of parameter space accessible to dark sunshine.

\subsection{Direct Detection Experiments}
Direct detection experiments bound the dark matter--nucleon cross section for weak scale dark matter, which in this model is a function of the dark coupling constant, the dark matter and dark photon mass, and the kinetic mixing parameter. The strongest current bounds come from the PANDAX-II experiment \cite{Cui:2017nnn}. In the inelastic dark matter framework, direct detection bounds are modified and generically produce weaker exclusion limits \cite{TuckerSmith:2001hy,TuckerSmith:2004jv,Blennow:2015hzp}.

Before we proceed it is a good idea to have a general idea of the orders of magnitude involved. In order for the lighter state to scatter at all against a nucleus of mass $m_N$ it must have a lab frame speed $w_\text{min}$ of
\begin{equation}
w_\text{min} = \sqrt{2 \Delta \left(\frac{1}{m_N} + \frac{1}{m_X}\right)} \ .
\end{equation}
We can estimate from this the size of the mass splitting that will forbid scattering. If dark matter has a characteristic velocity of $220 ~\text{km/s}$ and a mass much higher than that of the nucleus, and we take $m_N \sim 100 ~\gev$ we see that $\Delta \sim 100 ~\kev$. We now make this more precise.

The differential nuclear recoil rate in a detector whose fiducial volume encompasses $N_T$ nuclei with mass $m_N$, mass number $A$, and atomic number $Z$, when the dark matter velocities follow a Maxwell--Boltzmann distribution, is given by \cite{Freese:1987wu}
\begin{align}
\frac{dR}{dE_R} = \frac{N_T m_N \rho_X}{4 u_0 m_X} F_N^2(E_R) \frac{\sigma_{Xn}}{\mu_n^2} \frac{\left(f_p Z+ f_n(A-Z)\right)^2}{f_n^2} \left(\frac{\text{erf}(y_\text{min} + \eta) - \text{erf}(y_\text{min}-\eta)}{\eta}\right) \ .
\label{eq:DDdiffrate}
\end{align}
This formula depends on 
\begin{itemize}
	\item The Helm form factor: 
	\begin{align}
	F_N^2 &= \exp \left(- \frac{E_R}{E_N} \right)
	&
	E_N &= \frac{0.114~\gev}{A_N^{5/3}}
	\ .
	\end{align}
	\item The Earth's velocity in the galactic rest frame in units of the velocity of the local standard of rest, $u_0$:
	\begin{align}
	\eta \equiv \frac{u_\oplus}{u_0} = \frac{V_\odot + V_\oplus \cos \gamma \cos (\omega(t-t_0))}{u_0} \ ,
	\end{align}
	where $u_0 = 220 ~\text{km/s}$, $V_\odot = u_0+12 ~\text{km/s}$, $V_\oplus = 30 ~\text{km/s}$, $\omega = 2\pi/\text{yr}$, $t_0 = ~\text{June 2}$, and $\cos \gamma = 0.51$ describes the inclination of the Earth's orbital plane relative to the Sun's orbit about the galactic center. Both here and in our later examination of the dark matter capture rate we adhere to the conventions established by Gould \cite{Gould:1987ir,Gould:1987ju,Gould:1991hx,Gould:1999je}: $u$ denotes ``asymptotic'' velocities far from the gravitational influence of the Sun, $v$ denotes escape velocities, and $w$ denotes lab frame velocities at the point of interaction.
	\item The speed $y$ of a WIMP incident on the detector in units of $u_0$. Its lower limit $y_\text{min}$ corresponds to the lowest speed the WIMP can have while imparting a recoil energy $E_R$:
	\begin{align}
	y_\text{min} \equiv \frac{1}{u_0} \sqrt{\frac{1}{2 m_N E_R}} \left(\frac{m_N E_R}{\mu_N} + \Delta\right) \ ,
	\end{align}
	where $\mu_N$ is the reduced mass of the WIMP--nucleus system. This is in contrast to $\mu_n$, which denotes the reduced mass of the WIMP--nucleon system.
\end{itemize} 

We note that direct detection limits are usually derived with the assumption that the dark matter velocities in the galaxy follow a Maxwell--Boltzmann distribution, in contrast to our treatment of dark matter capture. We confirmed in previous work that the capture rate in the Sun is only weakly sensitive to this discrepancy and we therefore expect our results to be valid over all distributions favored by dark matter halo observations and simulations. 

\begin{figure}[t] 
	\hspace*{-.5cm} 
	\includegraphics[width=0.45\linewidth]{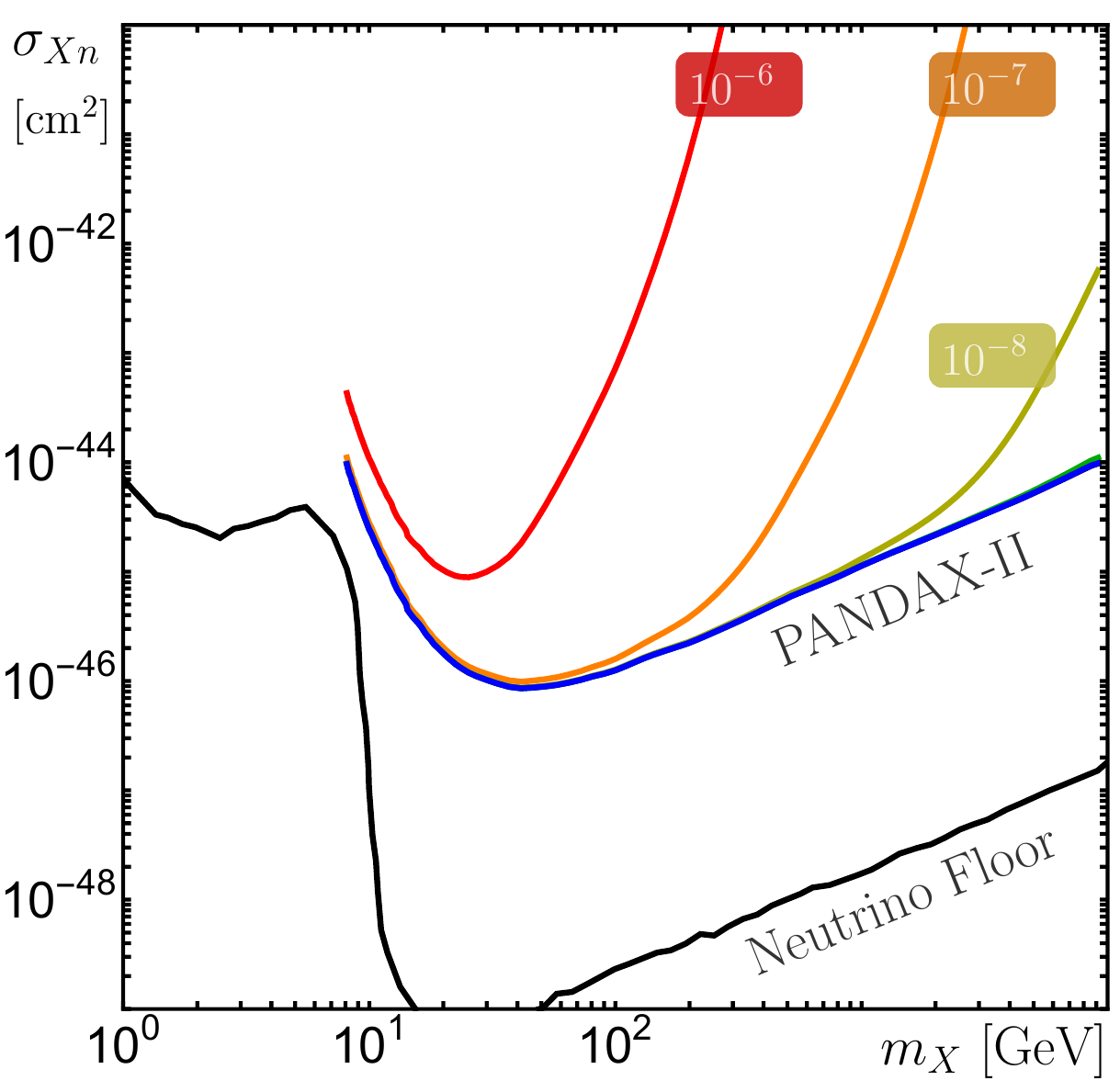} \qquad
	\includegraphics[width=0.45\linewidth]{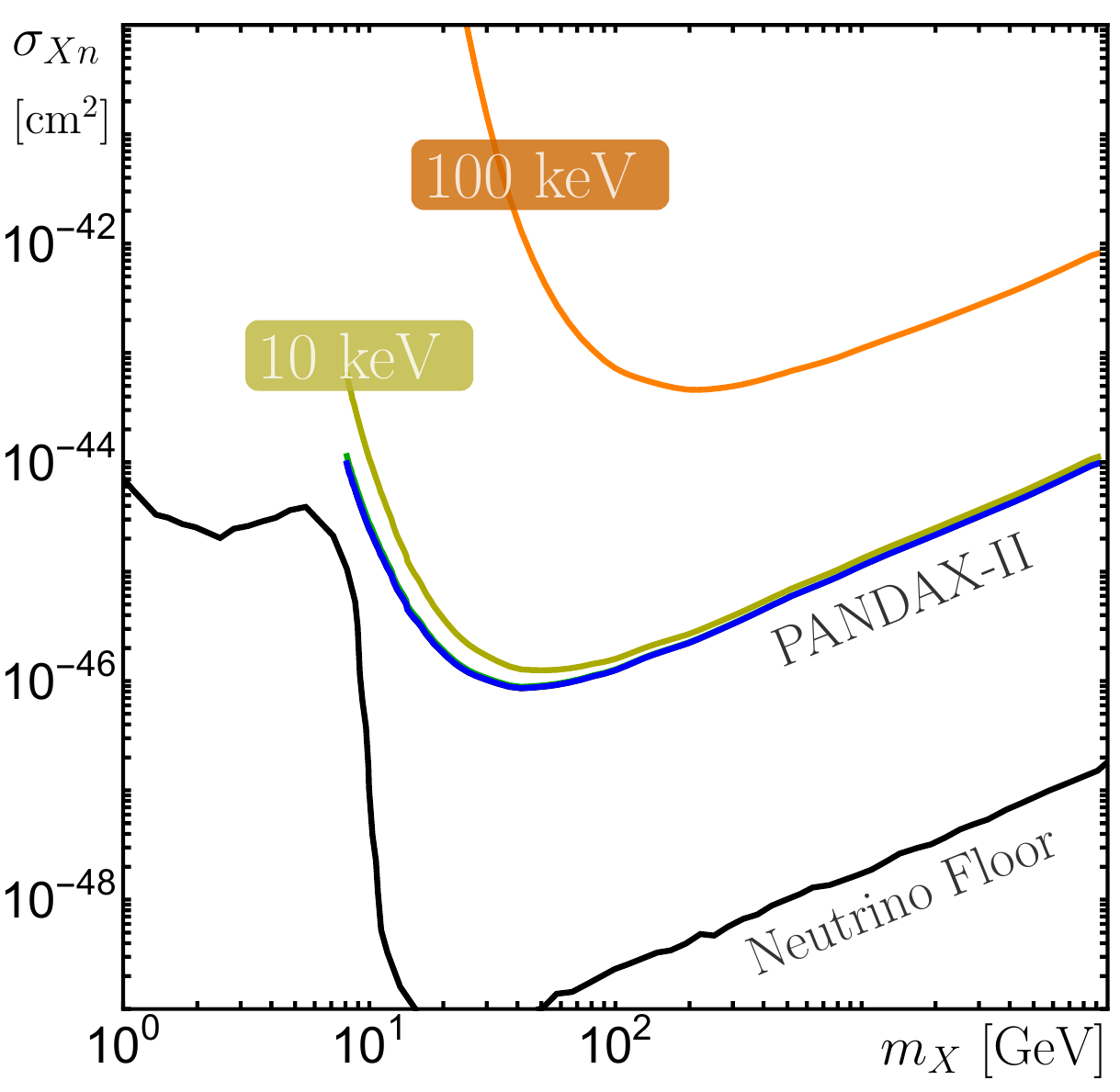} \\
	\vspace*{-0.1in} 
	\caption{ \textbf{Left:} Rescaling of PANDAX-II limits (\textsc{Blue}) \cite{Cui:2017nnn} on spin independent WIMP--nucleon scattering for indicated fixed values of $\delta \equiv \Delta/m_X$. \textsc{Red:} $\delta = 10^{-6}$, \textsc{Orange:} $\delta = 10^{-7}$, \textsc{Yellow:} $\delta = 10^{-8}$. \textbf{Right:} Same as Left, but for fixed values of $\Delta$. \textsc{Orange:} $\Delta = 100~\kev$. \textsc{Yellow:} $\Delta = 10 ~\kev$. Also shown is the neutrino floor, where coherent scattering of solar and atmospheric neutrinos will appear as background in direct detection experiments.
	}
	\label{fig:ddplaneinelastic}
	\vspace*{-0.1in}
\end{figure}

Over a given time interval, the expected number of events seen within a window of nuclear recoil energies is given by
\begin{align}
\frac{N_T m_N \rho_X}{4 u_0 m_X} \frac{\sigma_{Xn}}{\mu_n^2} \frac{\left(f_p Z+ f_n(A-Z)\right)^2}{f_n^2} \int d E_R ~d t ~F_N^2(E_R) \left(\frac{\text{erf}(y_\text{min} + \eta) - \text{erf}(y_\text{min}-\eta)}{\eta}\right) \ ,
\end{align}
where the limits of integration are determined by the experiment's nuclear recoil sensitivity and the live time of the experiment. In order to roughly evaluate the effect of a nonzero mass splitting on the PANDAX-II bounds, we notice that only the integral over recoil energy and time is dependent on $\Delta$, through the dependence on $y_\text{min}$. For a fixed detector composition and size we see that for each dark matter mass, upper limits on $\sigma_{Xn}$ scale as
\begin{equation}
\sigma^\text{upper}_{Xn} \propto \left[\int d E_R ~d t ~F^2(E_R) \left(\frac{\text{erf}(y_\text{min}(\Delta) + \eta) - \text{erf}(y_\text{min}(\Delta)-\eta)}{\eta}\right)\right]^{-1} \ .
\end{equation}
To explicitly evaluate the recoil integral we assume that the fiducial volume of PANDAX-II is composed entirely of nuclei of $Z = 54$ and $A= 131$. We assume that PANDAX-II has a nuclear recoil threshold of $1~\kev$, and average the Earth's velocity relative to the WIMP wind over a full annual cycle. The results of this scaling on the $(m_X,\sigma_{Xn})$ plane are shown in Figure \ref{fig:ddplaneinelastic}.

These bounds on the cross section can be translated into bounds on light mediators between the dark and visible sector \cite{DelNobile:2015uua}, and can therefore be shown on the dark photon $(m_{A'},\varepsilon)$ plane. To see how these bounds change when a dark sector mass splitting is introduced, we notice that the WIMP--nucleon cross section is quadratically dependent on the kinetic mixing parameter, and therefore that bounds on $\varepsilon$ will scale like the recoil integral raised to the $-1/2$ power. Representative effects of this scaling will be shown in \secref{results}. 

We caution the reader not to take such a scaling argument too literally. The imposition of a hard cutoff for the nuclear recoil threshold and simple integration over detector livetime ignores details of the PANDAX-II detector response and data analysis as well as the Poissonian nature of direct detection event rates, but this will suffice to give a rough understanding of how direct detection bounds are modified in inelastic models.

\subsection{Fixed Target Experiments}
Fixed target experiments attempt to produce dark photons by bremsstrahlung off of SM beam constituents and search for the decay of these dark photons to SM particles after they have propagated a large distance through shielding material. Because we assume that the dark photon can decay only to SM states, these searches are insensitive to the matter content and coupling of the dark sector. As such we are able to plot the bounds from fixed target dark photon searches in the $(m_{A'},\varepsilon)$ plane, and these bounds will remain unchanged as we sweep over $m_X$ and $\Delta$. The bounds that most overlap with our proposed search region for dark sunshine come from E137 \cite{Bjorken:1988as,Bjorken:2009mm} and LSND \cite{Batell:2009di,Essig:2010gu,PhysRevC.58.2489}, which jointly constrain dark photons with $1 ~\mev \lesssim m_{A'} \lesssim 300 ~\mev$ and $10^{-8} \lesssim \varepsilon \lesssim 10^{-5}$. For a review of fixed target dark photon searches see Ref.~\cite{Essig:2013lka}.

\subsection{Supernova Observations}
Light, weakly interacting particles may be emitted from core-collapse supernovae, shortening the resulting burst of energetic neutrinos. Observations of SN1987A constrain the couplings and masses of dark photons for $1 ~\mev \lesssim m_{A'} \lesssim 100 ~\mev$ and $10^{-9} \lesssim \varepsilon \lesssim 10^{-7}$ \cite{Dent:2012mx,Dreiner:2013mua,Kazanas:2014mca,Rrapaj:2015wgs,Mahoney:2017jqk}. These bounds, however, have recently come under scrutiny \cite{Rrapaj:2015wgs,Chang:2016ntp,Hardy:2016kme,Mahoney:2017jqk} with special focus placed on the production of dark photons through nucleus--nucleus bremsstrahlung and accounting for effects of the stellar environment. It is always prudent to claim the most conservative exclusion region, and for these purposes we use the exclusions provided by Ref.~\cite{Mahoney:2017jqk}. Another set of bounds can be placed from the absence of an $\mev$ $\gamma$ signal from SN1987A \cite{Kazanas:2014mca}, though these may also be affected by the treatment of nucleus--nucleus bremsstrahlung in supernovae. In the absence of a more complete published analysis we will continue to show these bounds, which apply for $1 ~\mev \lesssim m_{A'} \lesssim 100 ~\mev$ and $10^{-11} \lesssim \varepsilon \lesssim 10^{-9}$. 
The bounds from Ref.~\cite{Dreiner:2013mua} require that the dark matter is light enough ($m_X \lesssim 100 ~\mev$) to be produced in the supernova core and are thus not applicable here.

\subsection{Indirect Detection}
Observations of the diffuse positron spectrum constrain the dark sector coupling $\alpha_X$ \cite{Batell:2009zp,Schuster:2009au,Schuster:2009fc}. These constraints do not conflict with the thermal value of $\alpha_X$ that we use throughout this manuscript.

\subsection{Cosmology}
Cosmic microwave background observations constrain dark matter annihilation in the early universe and thus $\alpha_X$ \cite{Adams:1998nr,Chen:2003gz,Padmanabhan:2005es,Slatyer:2015jla}. Ref. \cite{Slatyer:2015jla} provides the most stringent constraints, but these bounds do not reach the thermal relic value of $\alpha_X$ for the masses we consider. Big bang nucleosynthesis also constrains properties of long-lived dark photons independently of the dark matter characteristics \cite{Fradette:2014sza}. The relevant bounds to our proposed search come from the observed abundance of $^4\text{He}$ and exclude dark photons of mass $300 ~\mev \lesssim m_{A'} \lesssim 10 ~\gev$ and kinetic mixing $10^{-12} \lesssim \varepsilon \lesssim 10^{-10}$.

\section{Dark Matter in the Sun}
Dark matter becomes captured by the Sun if scattering with nuclei results in the final state particle traveling too slowly to escape the Sun's gravity. The captured dark matter sinks to the core and reaches thermal equilibrium with the surrounding matter. As it accumulates in the core the dark matter annihilates into pairs of dark photons. Due to the low temperature of the Sun's core, the tree level annihilation process receives large corrections from Sommerfeld enhancement. A detailed treatment of this process is given in Ref.~\cite{Feng:2016ijc}. In this manuscript we will highlight the modifications to the calculation resulting from the presence of a mass splitting. We find that the main difference occurs in the capture process, while for the mass splittings we consider the annihilation is unchanged.

For the masses we consider it suffices to examine only capture from dark matter scattering off of nuclei, ignoring the effects of dark matter self-capture \cite{Zentner:2009is} and dark matter evaporation \cite{Griest:1986yu,Gaisser:1986ha}. The rate of dark matter annihilation in the Sun is
\begin{equation}
\Gamma_\text{ann} %= \frac{1}{2} N_X^2 C_\text{ann} 
= \frac{1}{2} C_\text{cap} \tanh^2 \frac{t}{\tau} \ ,
\end{equation}
where $C_\text{cap}$ is the rate of dark matter capture and $\tau$ is the timescale at which the capture and annihilation rates balance so that the total number of dark matter particles in the Sun is roughly constant. 
As we can see, when the lifetime of the body $t$ is less than the equilibrium timescale $\tau$ the annihilation rate drops quickly, while for $t > \tau$ the annihilation rate is approximately constant in time and proportional to the capture rate. The primary impact to our signal rate due to the addition of inelasticity stems from the capture kinematics, to which we now turn.

\subsection{Dark Matter Capture}
The differential rate for dark matter scattering off of nuclei is proportional to the densities of the interacting particles and their velocity distributions. We work in heliocentric coordinates in the Sun's rest frame, and take the nuclei comprising the Sun to be at rest. The differential rate is then
\begin{equation}
d^3r ~d^3 w ~dE_R ~n_X ~n_N(r) ~w ~f_\odot(w,r) \frac{d\sigma_N}{dE_R} \ ,
\end{equation}
where $w$ is the velocity of the incident dark matter, $r$ is the position, $n_X$ and $n_N$ denote the number densities of dark matter and nuclei, respectively, $f_\odot(w,r)$ is the velocity distribution of dark matter, and $d \sigma_N/dE_R$ is the differential scattering cross section. Dark matter passing through the solar system will become captured when it scatters off of a nucleus such that the final state dark sector particle has a velocity below the Sun's escape velocity. This requirement imposes a limit on the recoil energies leading to capture. We integrate the above rate over the volume of the body and over the appropriate limits on recoil energy and velocity to find the capture rate for a single species of nucleus $N$:
\begin{equation}
C_\text{cap}^N = n_X \int_0^{R} dr ~4\pi r^2 n_N(r) \int_{w_\text{min}}^{\infty} dw ~4\pi w^3 f_\odot(w,r) \int_\text{cap} dE_R ~\frac{d\sigma_N}{dE_R} \ .
\end{equation}
The nontrivial lower limit on the lab frame velocity integral derives from kinematics: in the model provided in \secref{model} the only allowed hidden sector vertex involves a transition from the lighter state to the heavier state or vice versa. If the incident dark matter's kinetic energy is too low to produce the heavier state, scattering simply does not occur. The threshold energy occurs when the heavier state and the nucleus are both at rest in the dark matter--nucleus CM frame after the collision, which implies
\begin{equation}
E_\text{thresh} = \Delta \left(\frac{m_X}{m_N} + 1\right)
\end{equation}
to leading order in $\Delta$. This lower limit on the incident dark matter's kinetic energy furnishes a lower limit on velocity, in the non-relativistic limit:
\begin{equation}
w_\text{min} = \sqrt{2 \Delta \left(\frac{1}{m_N} + \frac{1}{m_X}\right)} \ .
\end{equation}
The limits on recoil energy are determined in part by the capture requirement and in part by kinematics. The recoil energy is given in terms of the CM frame scattering angle $\theta_\text{CM}$ by 
\begin{equation}
E_R = \frac{\mu_N^2 w^2}{m_N} \left(1 - \cos \theta_\text{CM}\right) \ .
\end{equation}
The upper limit of $E_R$ then occurs when $\cos\theta_\text{CM} = -1$:
\begin{equation}
E_R^\text{max} = \frac{2 \mu_N^2 w^2}{m_N} \ .
\end{equation}
The lower limit is set by the requirement that the nucleus carry away enough of the incident energy that the outgoing WIMP is moving too slowly to escape the Sun. Conservation of energy yields
\begin{equation}
E_R^\text{min} = \frac{1}{2} m_X u^2 - \frac{1}{2} \Delta v_\odot^2(r) \ ,
\end{equation}
where $v_\odot$ is the escape velocity of the Sun and $u$ is the asymptotic velocity of the dark matter in the Sun's rest frame. This lower limit represents a cut on the final state phase space of the scattering process, as distinguished from the earlier discussion of the threshold velocity, which determines whether the reaction can occur.

Using the same differential cross section, dark matter velocity distribution \cite{Baratella:2013fya}, and solar model \cite{Serenelli:2009ww,Serenelli:2009yc,SerenelliWeb} as in Ref.~\cite{Feng:2016ijc}, the dark matter capture rate can be written compactly:
\begin{align}
C_\text{cap}
&= 32\pi^3 \varepsilon^2 \alpha_X \alpha n_X
\sum_N \frac{Z_N^2}{m_N E_N} \exp\left(\frac{m_{A'}^2}{2m_N E_N}\right)c_\text{cap}^N
\\
c_\text{cap}^N
&=
\int_0^{R_\odot}  dr \, r^2 n_N(r)
\int_{\sqrt{w_\text{min}^2 - v_\odot^2(r)}}^\infty du \, u f_\odot(u) 
\Theta(\Delta x_N)
\left[
\frac{e^{-x_N}}{x_N} + \text{Ei}(-x_N)
\right]^{x_N^\text{min}}_{x_N^\text{max}} \ .
\label{eq:DM:capture:rate:full}
\end{align}
We have used a substitution variable
\begin{equation}
x_N \equiv \frac{2 m_N E_R + m_{A'}^2}{2 m_N E_N} \ ,
\end{equation}
with $E_N$
\begin{equation}
E_N = \frac{0.114~\gev}{A_N^{5/3}} \ ,
\end{equation}
and $\text{Ei}$ is the exponential integral function defined by
\begin{equation}
\text{Ei}(z) \equiv -\int_{-z}^{\infty} dt \frac{e^{-t}}{t} \ .
\end{equation}
We take the number density of dark matter in the solar system to be $n_X = 0.3 ~\gev/\text{cm}^3/m_X$.

We have verified that the treatment of dark matter capture presented above predicts the same $\mathcal{O}(1)$ suppression of solar capture for $\Delta \sim 100 ~\kev$ relative to the elastic case as that presented in Ref.~\cite{Menon:2009qj}, in the contact limit where $m_{A'}$ is large and the differential cross section's dependence on $E_R$ enters only though the Helm form factor. For smaller values of $m_{A'}$ the capture rate is further suppressed.

In our earlier examination of ``dark earthshine'' at IceCube \cite{Feng:2015hja} we found that such a search can probe a large region of parameter space currently inaccessible to direct detection experiments. Given the modification of direct detection bounds in an inelastic framework, it is natural to ask if this virtue persists.
The answer to this question is negative. We find numerically that when the dark matter mass splitting is large enough to significantly weaken direct direction constraints, the capture rate for dark matter in the Earth falls as well. This is an obvious result, since the scattering process for dark matter capture in the Earth is identical to that exploited by direct detection experiments. When the mass splitting is large enough that nuclear recoils in direct detection experiments are kinematically disallowed, we expect the capture process to be forbidden as well. For the mass splittings treated here we find that dark matter is simply moving too slowly around the Earth to be captured efficiently. With smaller mass splittings the direct detection exclusions are not qualitatively weakened while the potential signal rate at IceCube remains unchanged. For this reason we omit further discussion of dark matter capture in the Earth from the ensuing analysis and treat only the case of dark matter accumulated in the Sun, detectable by dark sunshine signatures at AMS-02.

\subsection{Dark Matter Annihilation}
The captured dark matter population can be taken to be in thermal equilibrium with the solar core when the WIMP--proton spin independent scattering cross section is greater than $10^{-51} ~\cm^2$ for $m_X = 100 ~\gev$, $10^{-50} ~\cm^2$ for $m_X = 1 ~\tev$, or $10^{-47} ~\cm^2$ for $m_X = 10 ~\tev$ \cite{Peter:2009mm}. We found in previous work that the cross section is several orders of magnitude larger than this in the relevant regions of parameter space, so that our analysis of dark matter annihilation is consistent. However, there is another wrinkle introduced in the framework of inelastic dark matter that we have already mentioned: in a theory without an elastic vertex slowly traveling dark matter will not scatter, and therefore the dark matter population will not come to thermal equilibrium with the Sun's core if its temperature is lower than the dark matter mass splitting \cite{Blennow:2015hzp}. This obstacle can be circumvented, however, by the presence of a small elastic coupling with strength $g'_X$. In order for a large enough WIMP--proton SI cross section that allows thermalization we may take $g'_X \sim 10^{-2} g_X$, which is too small to affect our treatment of capture, annihilation, and direct detection constraints in the inelastic framework \cite{Nussinov:2009ft, Blennow:2015hzp}. Such a coupling arises naturally at the one-loop level from the bare Lagrangian provided in \secref{model}.

The annihilation coefficient, encoding the dependence of the captured dark matter annihilation rate on the spatial distribution and thermal cross section, is given by \cite{Baratella:2013fya}
\begin{equation}
C_{\text{ann}} = \langle \sigma_{\text{ann}} v \rangle 
\left[ \frac{G_N m_X \rho_\odot}{3 T_\odot} \right] ^{3/2} \ .
\end{equation}
Here $T_\odot = 1.5 \times 10^7 ~\text{K} \approx 1.3 ~\kev$ is the temperature at the center of the Sun, $\rho_\odot = 151 ~\text{g/cm}^3$ is the matter density at the core of the Sun, and $G_N$ is Newton's gravitational constant.

Because of the low temperature of the core of the Sun relative to the dark matter mass, captured and thermalized dark matter will move very slowly and the annihilation process is Sommerfeld enhanced \cite{Sommerfeld:1931}. This modifies the annihilation cross section from the tree level result to
\begin{equation}
\langle\sigma_\text{ann} v\rangle = \langle S\rangle \langle\sigma_\text{ann} v\rangle_\text{tree} \ ,
\end{equation}
where $\langle S \rangle$ is the thermally-averaged Sommerfeld enhancement, given by
\begin{equation}
\langle S \rangle = \int \frac{d^3 v}{(2 \pi v_0^2)^{3/2}} e^{-v^2/v_0^2} S \ ,
\end{equation}
with $v_0$
\begin{equation}
v_0 = \sqrt{\frac{2 T_\odot}{m_X}} \approx 5.1 \times 10^{-5} \sqrt{\frac{\tev}{m_X}} \ .
\end{equation}
Here $S$ is the Sommerfeld enhancement for a two-body system with definite relative velocity, which we define below.

A semi-analytic approximation for the s-wave Sommerfeld enhancement with a massive mediator coupling non-degenerate states can be found by solving the two-state Schr\"odinger equation with a matrix-valued potential encoding the mediator exchange diagrams \cite{Slatyer:2009vg}. 
The result of this analysis is
\begin{align}
S = \frac{\pi}{\epsilon_v} \sinh \left(\frac{\epsilon_v \pi}{\mu} \right) \begin{cases}
\frac{1}{\cosh(\epsilon_v \pi/\mu) - \cos\left(\sqrt{\epsilon_\delta^2 - \epsilon_v^2}\pi/\mu + 2 \theta_-\right)} & \epsilon_v < \epsilon_\delta \\
	\frac{\cosh\left(\left(\epsilon_v + \sqrt{-\epsilon_\delta^2 + \epsilon_v^2}\right)\pi/2\mu\right)\text{sech}\left(\left(\epsilon_v-\sqrt{-\epsilon_\delta^2 + \epsilon_v^2}\right)\pi/2\mu\right)}{\cosh\left(\left(\epsilon_v + \sqrt{-\epsilon_\delta^2 + \epsilon_v^2}\right)\pi/\mu\right)-\cos(2\theta_-)} & \epsilon_v > \epsilon_\delta
\end{cases} \ ,
\label{eq:inelsomm}
\end{align}
where $\epsilon_v = v/\alpha_X$, $\epsilon_\delta = \sqrt{2\Delta/m_X}/\alpha_X$, $\mu$ is given by
\begin{equation}
\mu = \epsilon_\phi \left(\frac{1}{2} + \frac{1}{2}\sqrt{1+\frac{4}{\epsilon_\phi r_M}}\right) \ ,
\end{equation}
with $\epsilon_\phi = m_{A'}/m_X \alpha_X$, and $r_M$ is defined implicitly by
\begin{equation}
\frac{e^{-\epsilon_\phi r_M}}{r_M} = \text{max}\left[\epsilon_\delta^2/2, \epsilon_\phi^2\right] \ .
\end{equation}
The quantity $\theta_-$ is
\begin{equation}
\theta_- = \frac{1}{i}\int_{r_S}^{r_M} \sqrt{\tilde{\lambda_-}} ~dr - 4 z_S^{1/4} - \frac{1}{i} \int_0^{r_M} \sqrt{\lambda_-}~dr
\end{equation}
where $\tilde{\lambda_-}$ is given by
\begin{equation}
\tilde{\lambda_-} = -\epsilon_v^2 + \epsilon_\delta^2/2 - \sqrt{(\epsilon_\delta^2/2)^2 + V_0^2 e^{-2 \mu r}} \ ,
\end{equation}
with $V_0 \equiv \exp\left[\epsilon_\phi r_M \left(\sqrt{1+ 4/\epsilon_\phi r_M}-1\right)/2\right]/r_M$, $z_S = V_0^2 e^{-2 \mu r_S}/16 \mu^4$ with $r_S$ chosen such that $V_0 e^{-\mu r_S} \gg \epsilon_v^2, \epsilon_\delta^2$, and with $\lambda_-$ defined
\begin{equation}
\lambda_- = -\epsilon_v^2 + \epsilon_\delta^2/2 - \sqrt{(\epsilon_\delta^2/2)^2 + e^{-2 \epsilon_\phi r}/r^2} \ .
\end{equation}

For the mass splittings we consider in this manuscript, with $\Delta/m_X \sim 10^{-9} - 10^{-6}$ this formula is well approximated by sending $\epsilon_\delta \rightarrow 0$, recovering the Sommerfeld factor from the elastic case.

\begin{figure}[t] 
	\hspace*{-.5cm} 
	\includegraphics[width=0.41\linewidth]{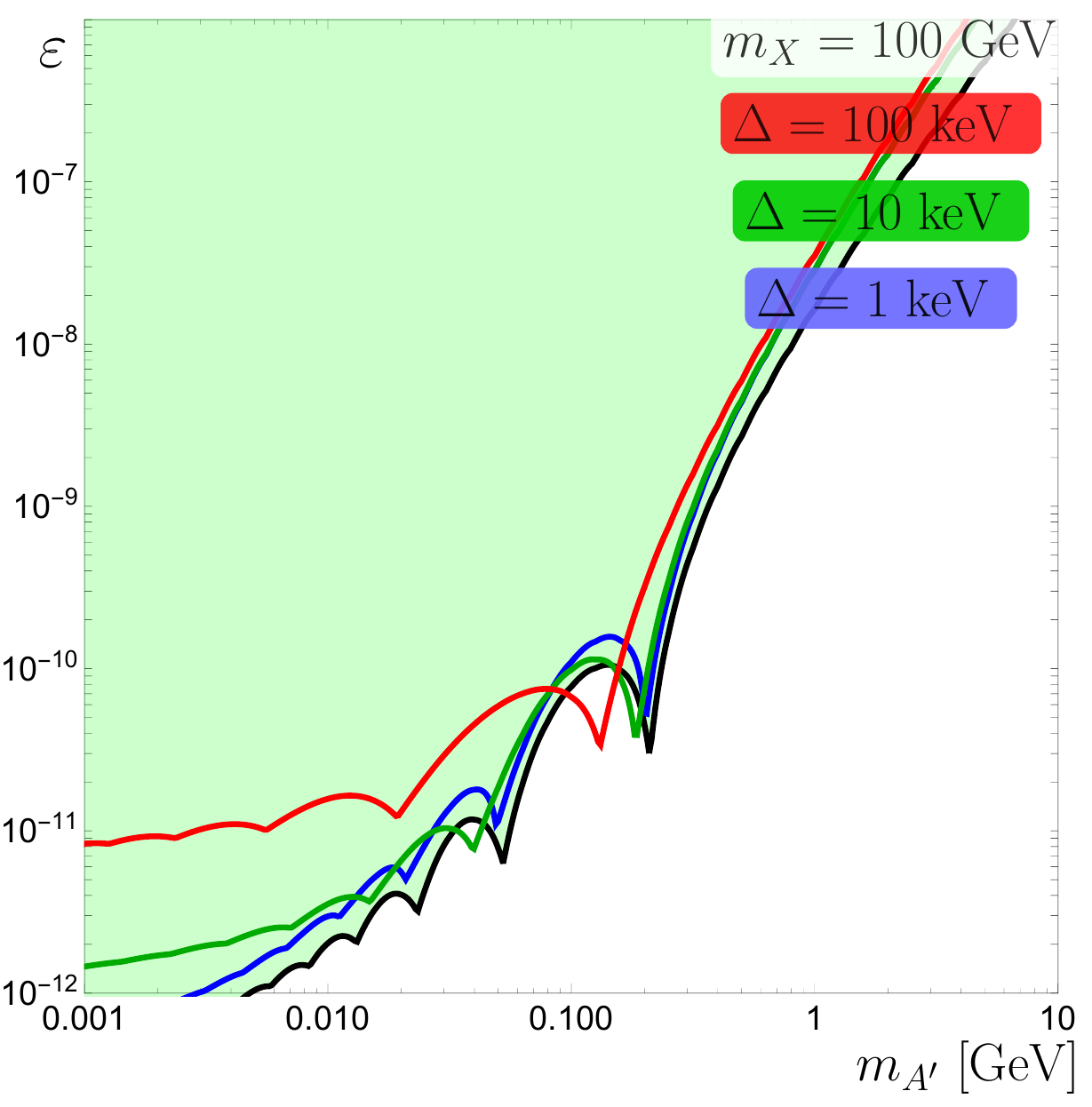} \qquad
	\includegraphics[width=0.41\linewidth]{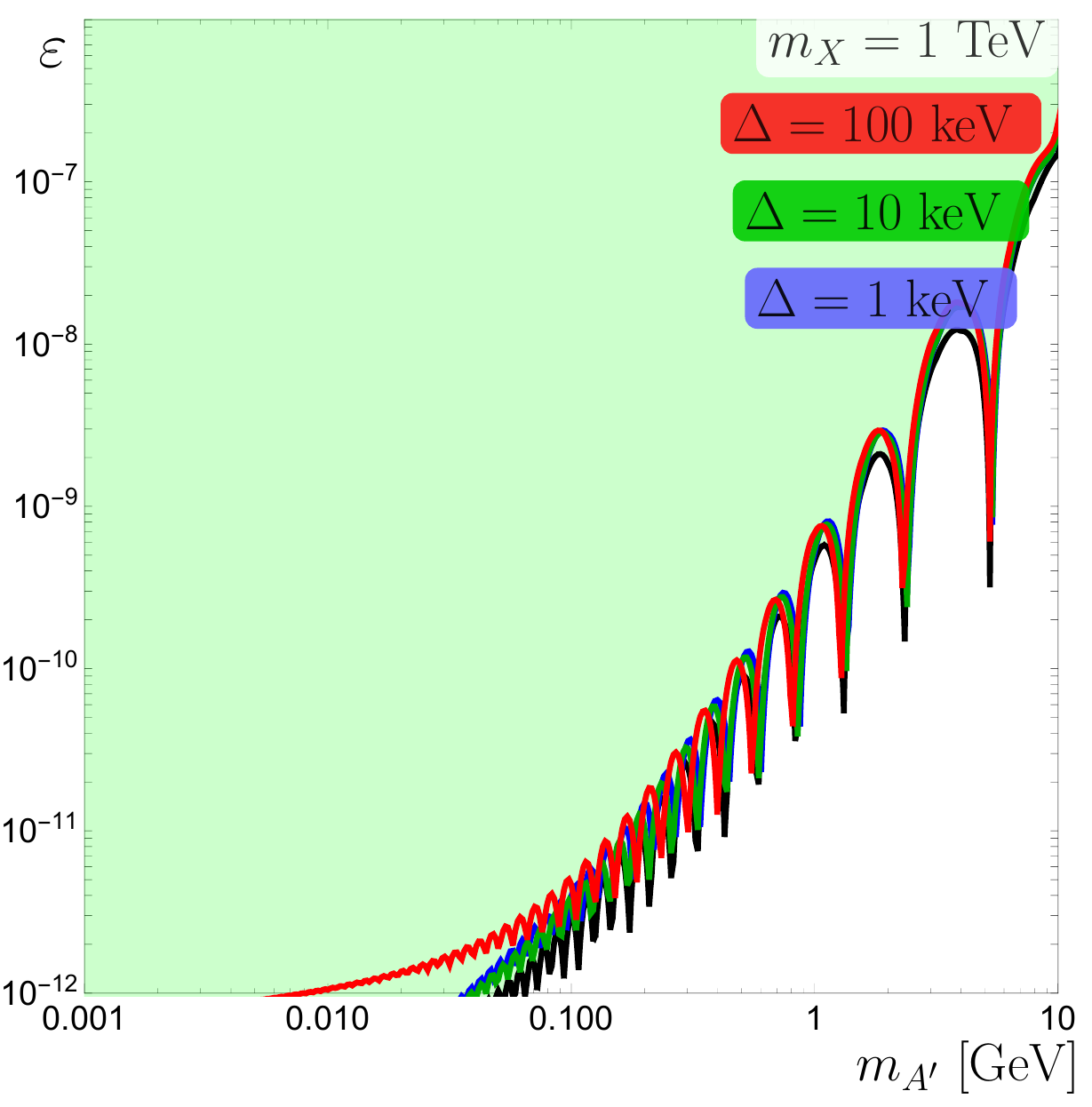} \\
	\includegraphics[width=0.41\linewidth]{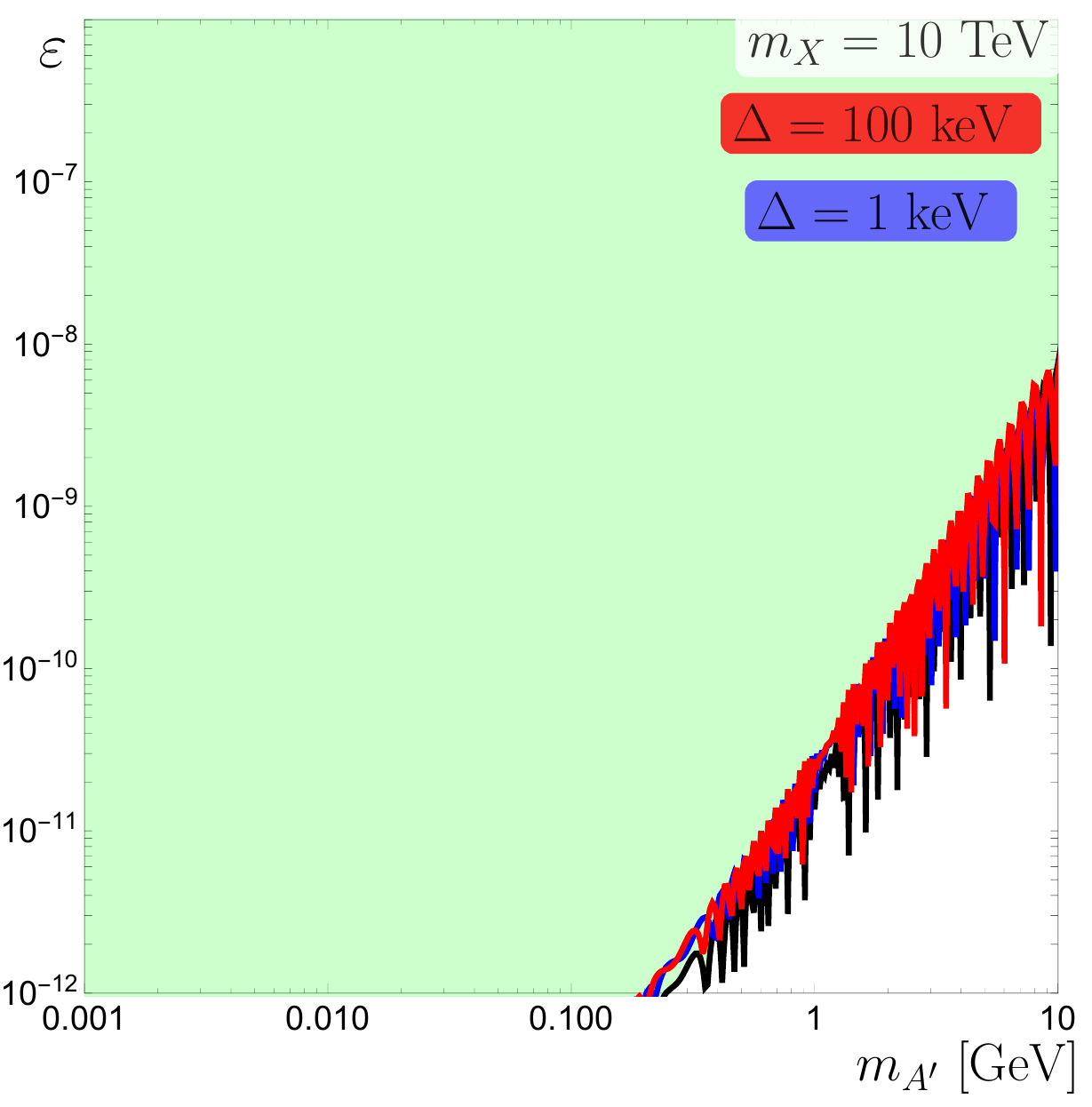} \qquad
	\vspace*{-0.1in} 
	\caption{Contours of $\tau/\tau_\odot = 1$ in the $(m_{A'},\varepsilon)$ plane for benchmark values of $m_X$ and $\Delta$, where $\tau_\odot= 4.6 \times 10^9 ~\text{yr}$ is the age of the Sun. \textbf{Black:} Contours of $\tau/\tau_\odot = 1$ for the elastic case $\Delta = 0$. \textbf{Red:} Same, but for $\Delta = 100 ~\kev$. \textbf{Green:} Same, but for $\Delta = 10 ~\kev$. \textbf{Blue:} Same, but for $\Delta = 1 ~\kev$. The dark sector coupling $\alpha_X$ is fixed by requiring that this dark sector match the observed density of dark matter as a thermal relic. For parameter values lying in the shaded region above and to the left of the given contours the captured dark matter population in the Sun is in equilibrium today.
	}
	\label{fig:Suntaus}
	\vspace*{-0.1in}
\end{figure}

\subsection{Equilibrium Time}
The equilibrium time $\tau = 1/\sqrt{C_\text{cap} C_\text{ann}}$ for the population of captured dark matter can be evaluated using the results from the two previous sections. We calculate the equilibrium time for the benchmark masses $m_X = 100 ~\gev$, $1 ~\tev$, and $10~\tev$ over a range of the dark photon parameter space, for the mass splittings $\Delta = 100$, $10$, and $1 ~\kev$. The regions of dark photon parameter space over which the Sun is currently in equilibrium are shown in \figref{Suntaus}. We see that at large masses the equilibrium times are very nearly the same as in the elastic case, and the reduction of the capture rate as the mass splitting increases is reflected most clearly for the $100 ~\gev$ case.

\section{Detection}
\label{sec:detection}

Dark matter annihilations will produce dark photons that travel outwards from the Sun. If those dark photons are massive enough, they will decay to produce highly boosted $e^+ e^-$ pairs. These energetic charged particles will be deflected by the Sun's magnetic field by the time they arrive at Earth. We suggest cuts on the energy and incidence direction of positrons so that the number of background positrons satisfying these cuts is reduced to $1$. These cuts are derived in Ref.~\cite{Feng:2016ijc}, and their effect is to include a multiplicative factor $P_\text{det}$ correcting a na{\"i}ve estimate of the flux of dark photons:
\begin{equation}
N_\text{sig} = 2 \Gamma_\text{ann} \frac{\xi_\odot}{4 \pi ~\text{au}^2} ~\text{Br}\left(A' \rightarrow e^+ e^-\right) P_\text{det} \ .
\end{equation}
The factor $\xi_\odot$ denotes the ``exposure'' and has dimensions of $\text{area} \times \text{time}$. Assuming uniform operating conditions and accounting for the fact that AMS-02 is only facing the Sun for a fraction of its livetime, its numerical value is \cite{Machate}
\begin{equation}
\xi_\odot = 6.3 \times 10^4 ~\text{m}^2\text{s} ~\frac{T}{\text{yr}} \ ,
\end{equation}
where $T$ is the total time over which AMS-02 has been operating in orbit.

Explicitly, $P_\text{det}$ is
\begin{equation}
P_\text{det} = \frac{m_X - E_\text{cut}}{m_X} \left(e^{-R_\odot/L} - e^{-\text{au}/L}\right) - P^B_\text{det} \Theta\left(E_0 \log\frac{\text{au}}{R_\odot} - E_\text{cut}\right) \ ,
\end{equation}
where $P^B_\text{det}$ is given by
\begin{equation}
P^B_\text{det} = \frac{E_0}{m_X} \left[\text{Ei}\left(-\frac{\text{au}}{L} e^{-E_\times/E_0}\right) - \text{Ei}\left(-\frac{\text{au}}{L}e^{-E_\text{cut}/E_0}\right)\right] + \frac{E_\times - E_\text{cut}}{m_X} e^{-R_\odot/L} \ .
\end{equation}
Here $E_0$ and $E_\times$ are defined
\begin{align}
E_0(E_\text{cut}) &\equiv 1.5 ~\tev \left(\frac{100 ~\gev}{E_\text{cut}}\right)^{0.9} \sqrt{\frac{T}{\text{yr}}} \ ,
&
E_\times &\equiv \text{min}\left(E_0 \log\frac{\text{au}}{R_\odot},m_X\right) \ .
\end{align}
At each point in our scan of parameter space we numerically maximize $P_\text{det}$ as a function of $E_\text{cut}$ with the minimal allowed $E^\text{min}_\text{cut} = 50 ~\gev$ corresponding to the energy threshold at AMS-02.

\begin{figure}[t] 
	\hspace*{-.5cm}
	\includegraphics[width=0.45\linewidth]{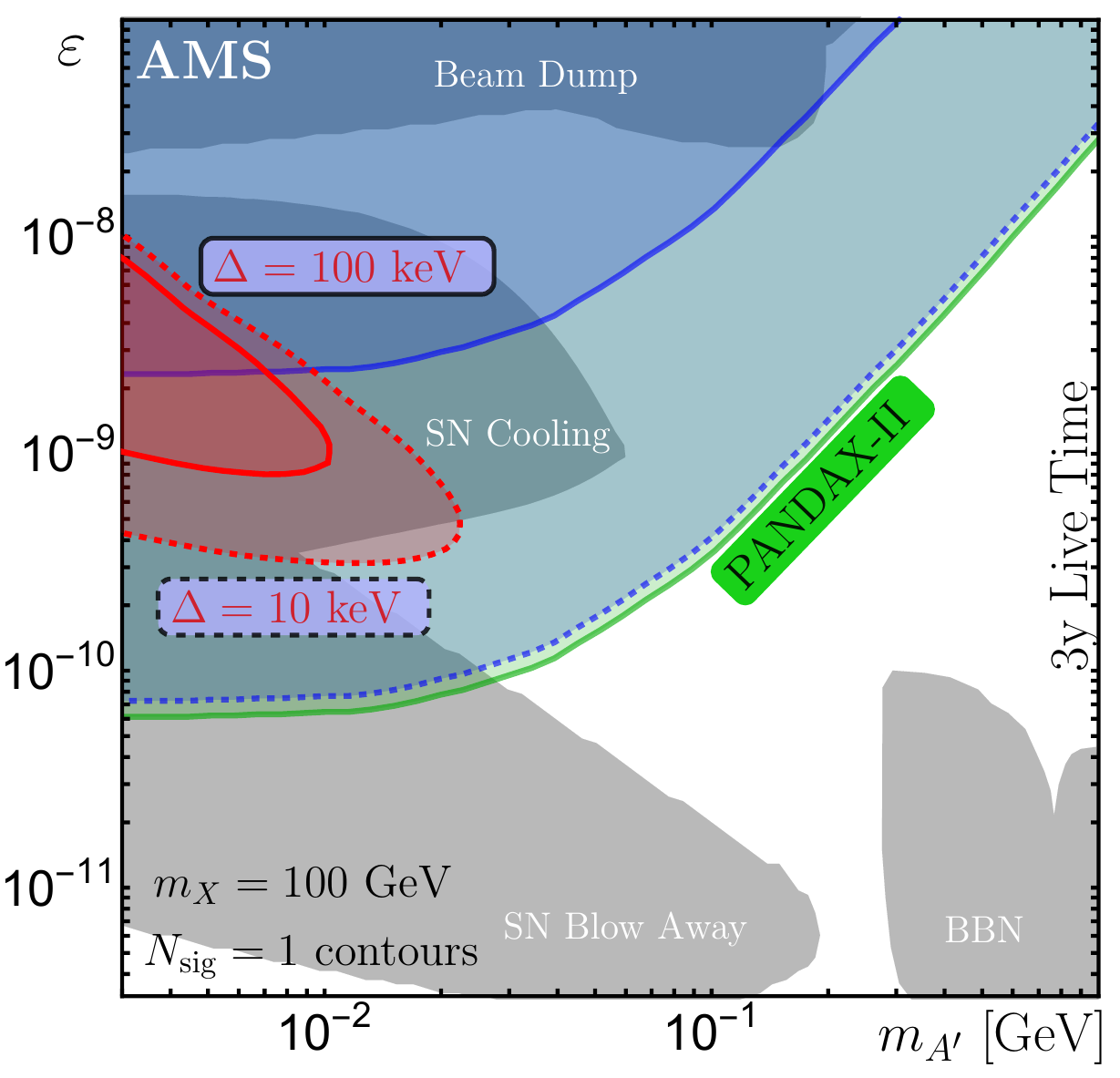} \qquad
	\includegraphics[width=0.45\linewidth]{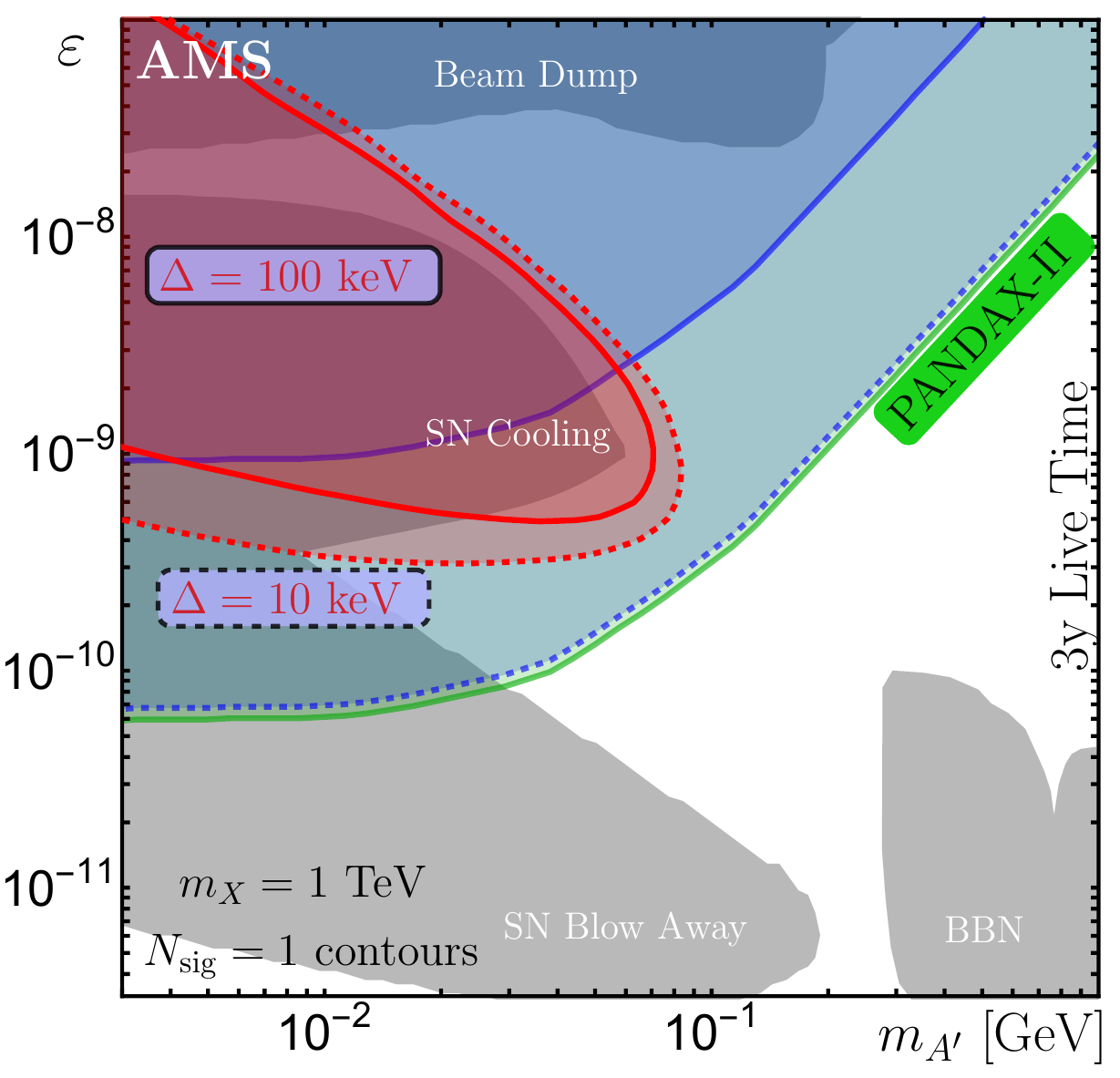} \\
	\includegraphics[width=0.45\linewidth]{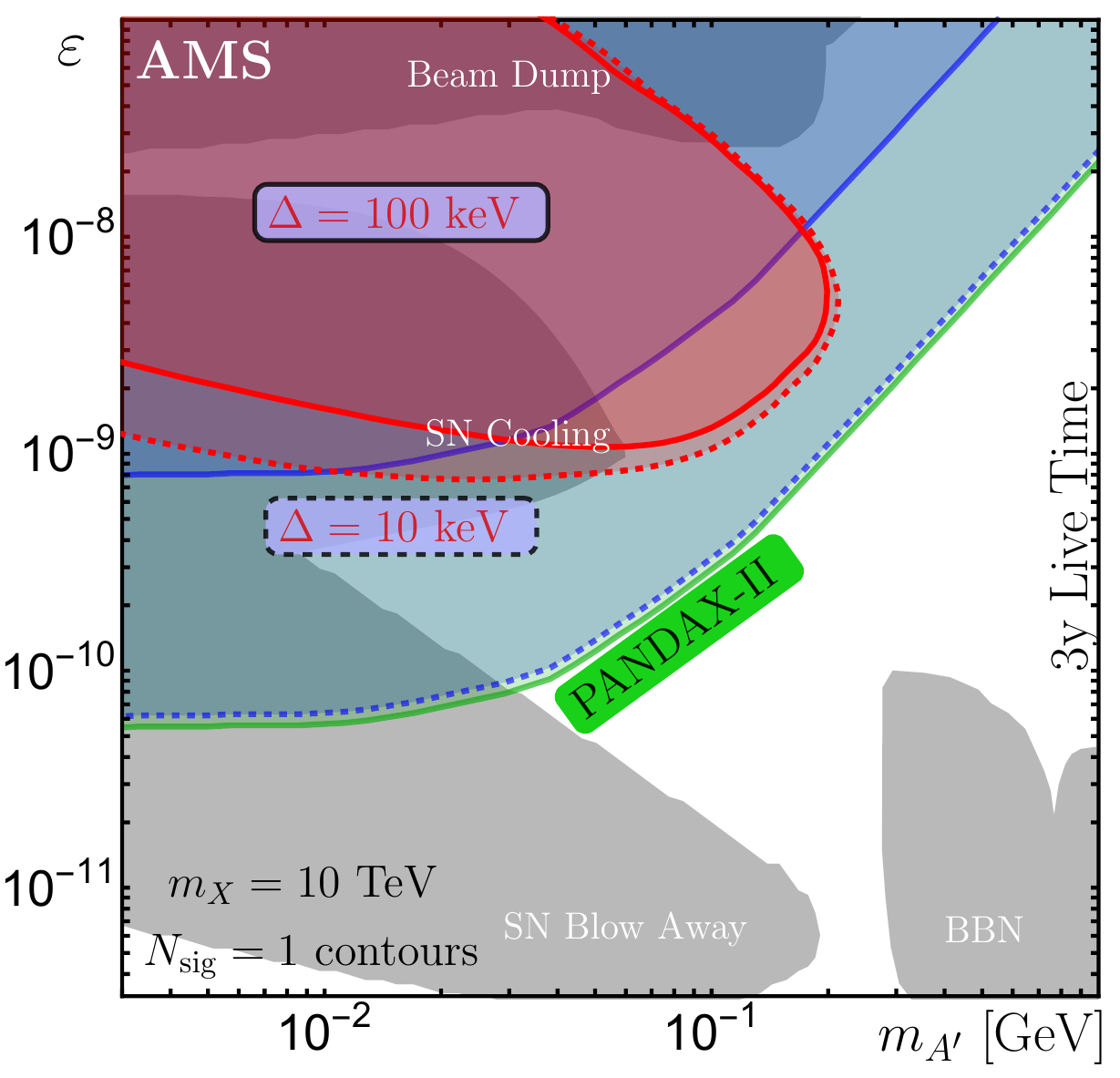} \qquad
	\includegraphics[width=0.45\linewidth]{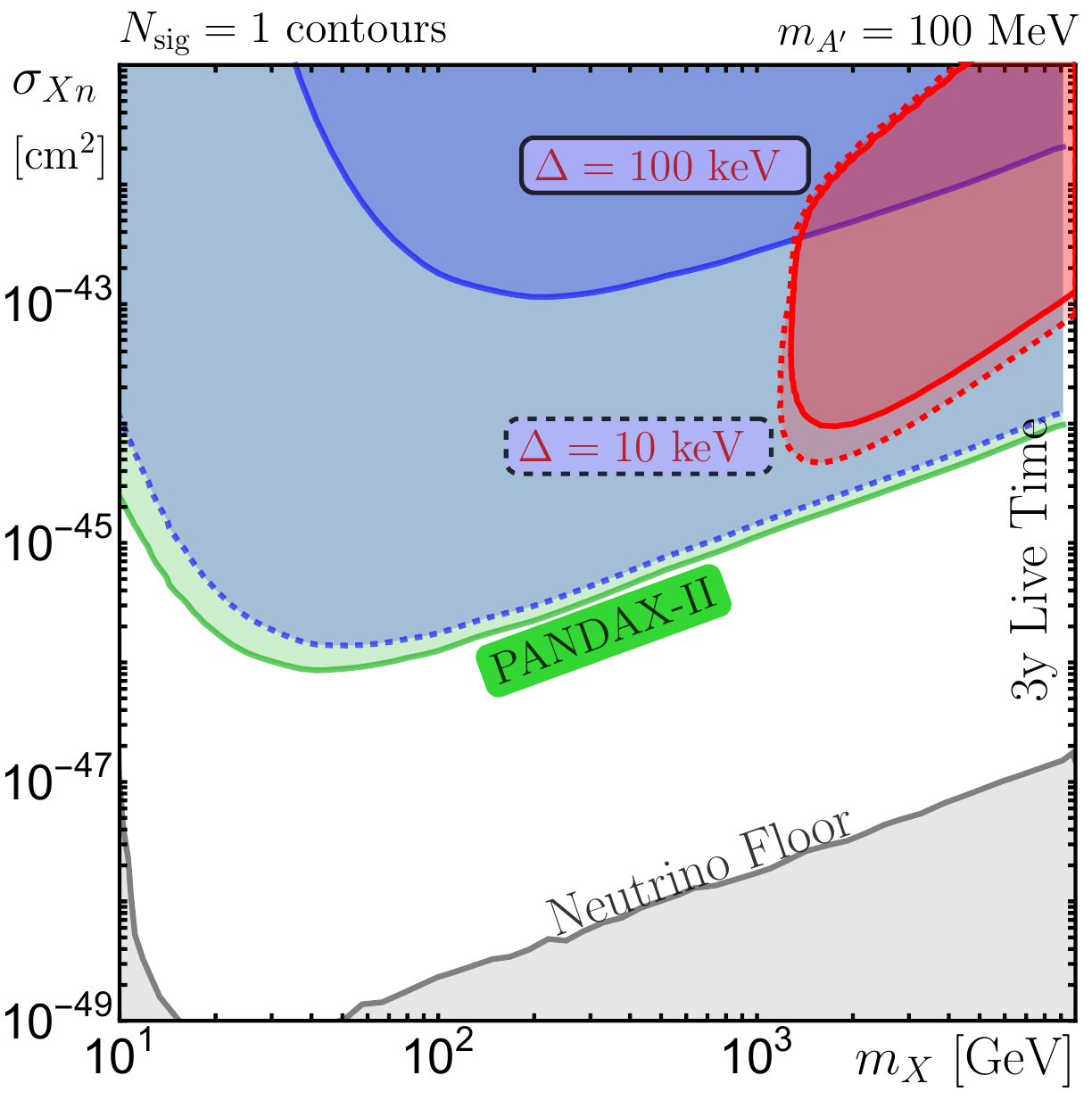}
	\vspace*{-0.1in} 
	\caption{\textbf{Red:} AMS-02 reach region for $T=3$ years live
		time in the $(m_{A'}, \varepsilon)$ plane for $m_X = 100~\gev$
		(top left), $m_X = 1~\tev$ (top right), and $m_X = 10~\tev$ (bottom
		left); and in the $(m_X,\sigma_{Xn})$ plane for $m_{A'} = 100 ~\mev$ (bottom right).  The dark sector fine-structure constant $\alpha_X$ is set
		by requiring $\Omega_X \simeq 0.23$. Solid curves are for the dark sector mass splitting $\Delta = 100 ~\kev$, while dashed curves are for $\Delta = 10 ~\kev$. The indirect detection reach is
		also compared to other probes.  \textbf{Green:} current bounds from
		direct detection with $\Delta = 0$ ~\cite{DelNobile:2015uua,Cui:2017nnn}. \textbf{Blue:} rescaled bounds from direct detection. Solid curves are for $\Delta = 100~\kev$, while dashed curves are for $\Delta=10~\kev$. \textbf{Gray:} regions
		probed by beam dump experiments, supernova observations, and BBN
		constraints~\cite{Essig:2013lka,Fradette:2014sza} (top and bottom left), and the neutrino floor (bottom right). }
	\label{fig:results}
	\vspace*{-0.1in}
\end{figure}

\begin{figure}[t] 
	\hspace*{-.5cm}
	\includegraphics[width=0.45\linewidth]{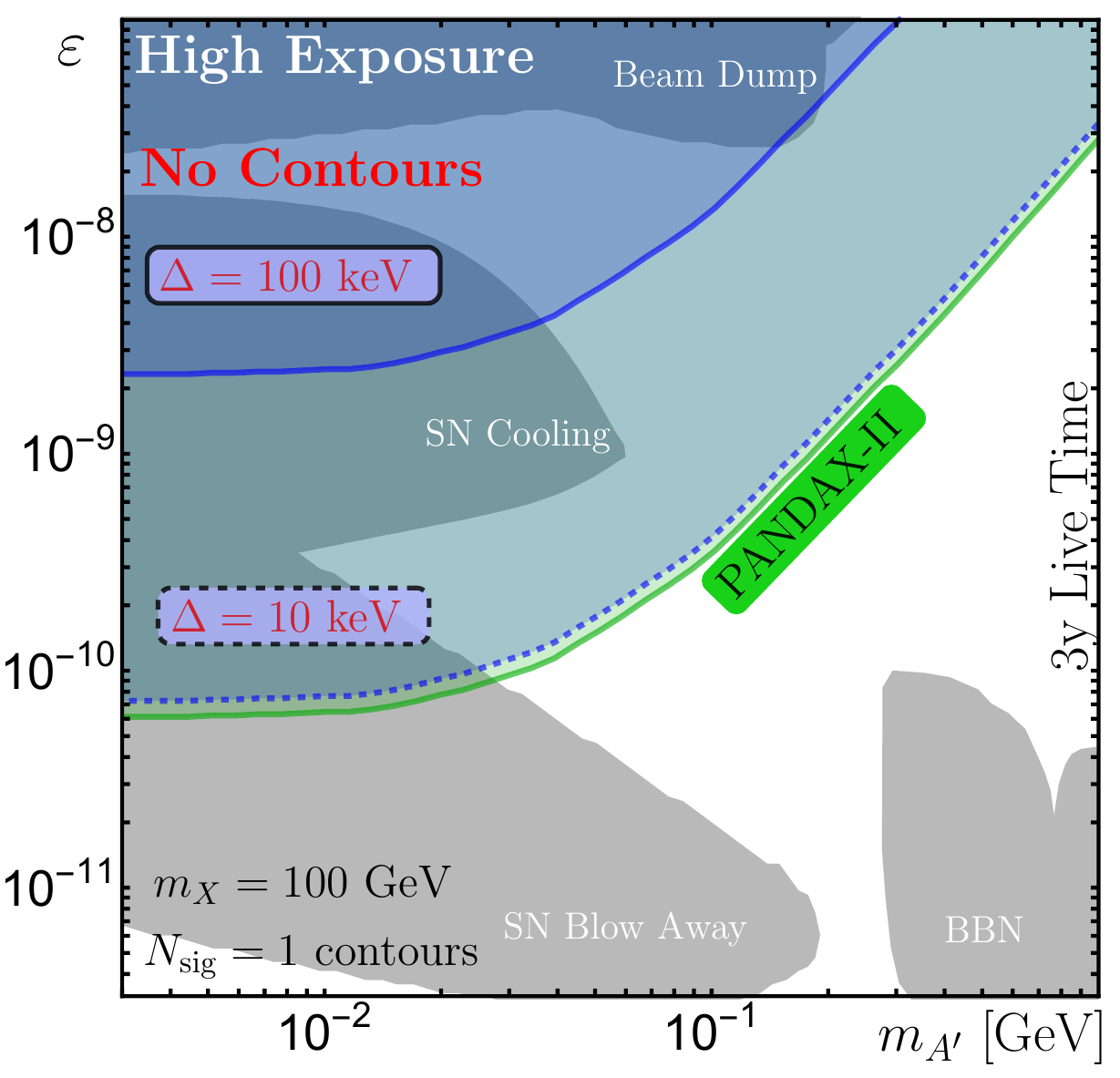} \qquad
	\includegraphics[width=0.45\linewidth]{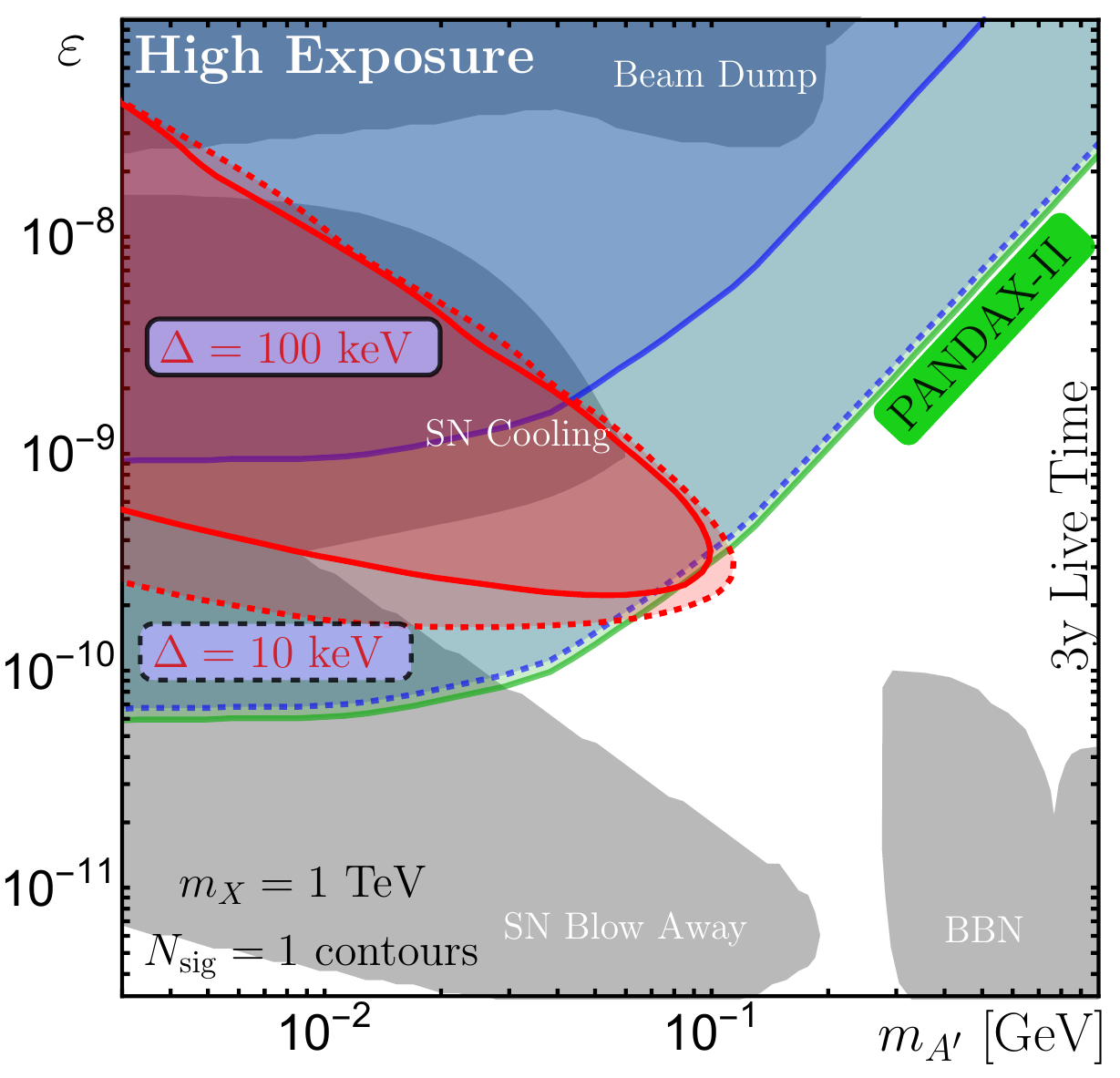} \\
	\includegraphics[width=0.45\linewidth]{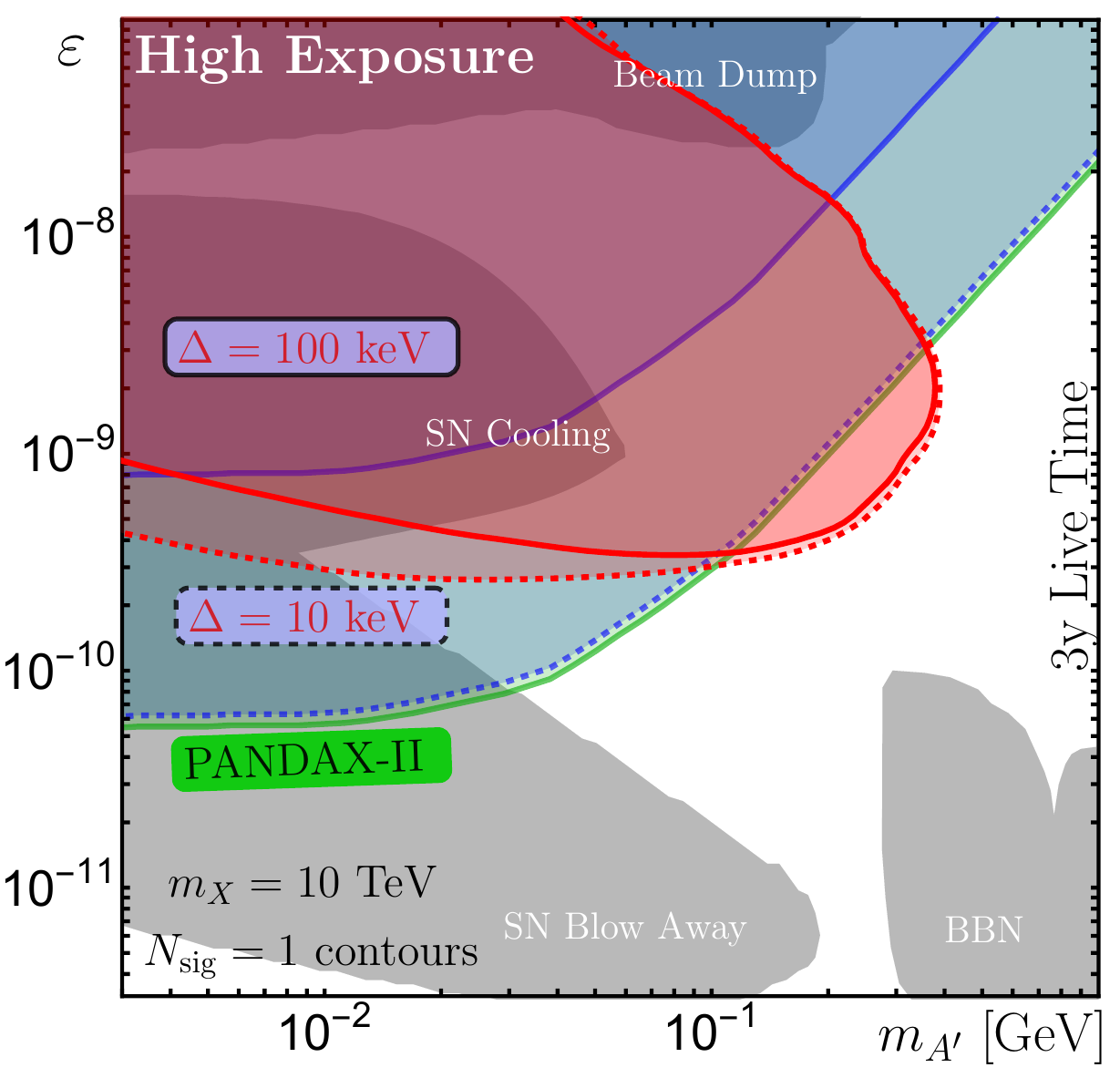} \qquad
	\includegraphics[width=0.45\linewidth]{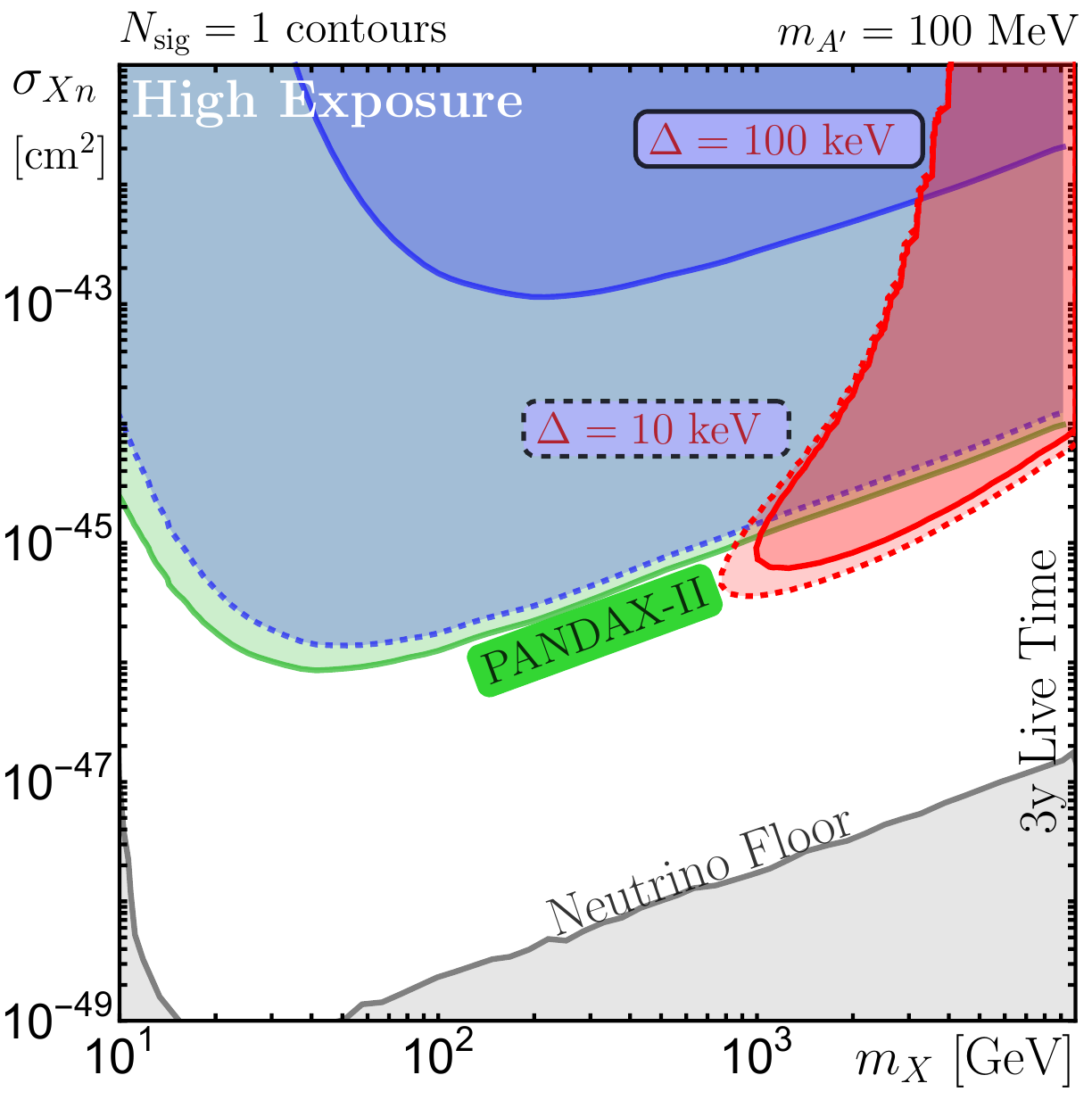}
	\vspace*{-0.1in} 
	\caption{Same as \figref{results} but with increased exposure $\xi_\odot^\text{high} = 80 \xi_\odot$. Note that the reach for $m_X = 100~\gev$ dark matter vanishes at AMS-02. This is because our angular and energy cuts, set by the condition that we expect only one background event, restrict the amount of signal positrons to be negligibly small. The positron background drops off at higher energies, so that a search for more massive dark matter does not suffer the same reduction in expected signal rate.}
	\label{fig:resultshighexposure}
	\vspace*{-0.1in}
\end{figure}

\begin{figure}[t] 
	\hspace*{-.5cm}
	\includegraphics[width=0.45\linewidth]{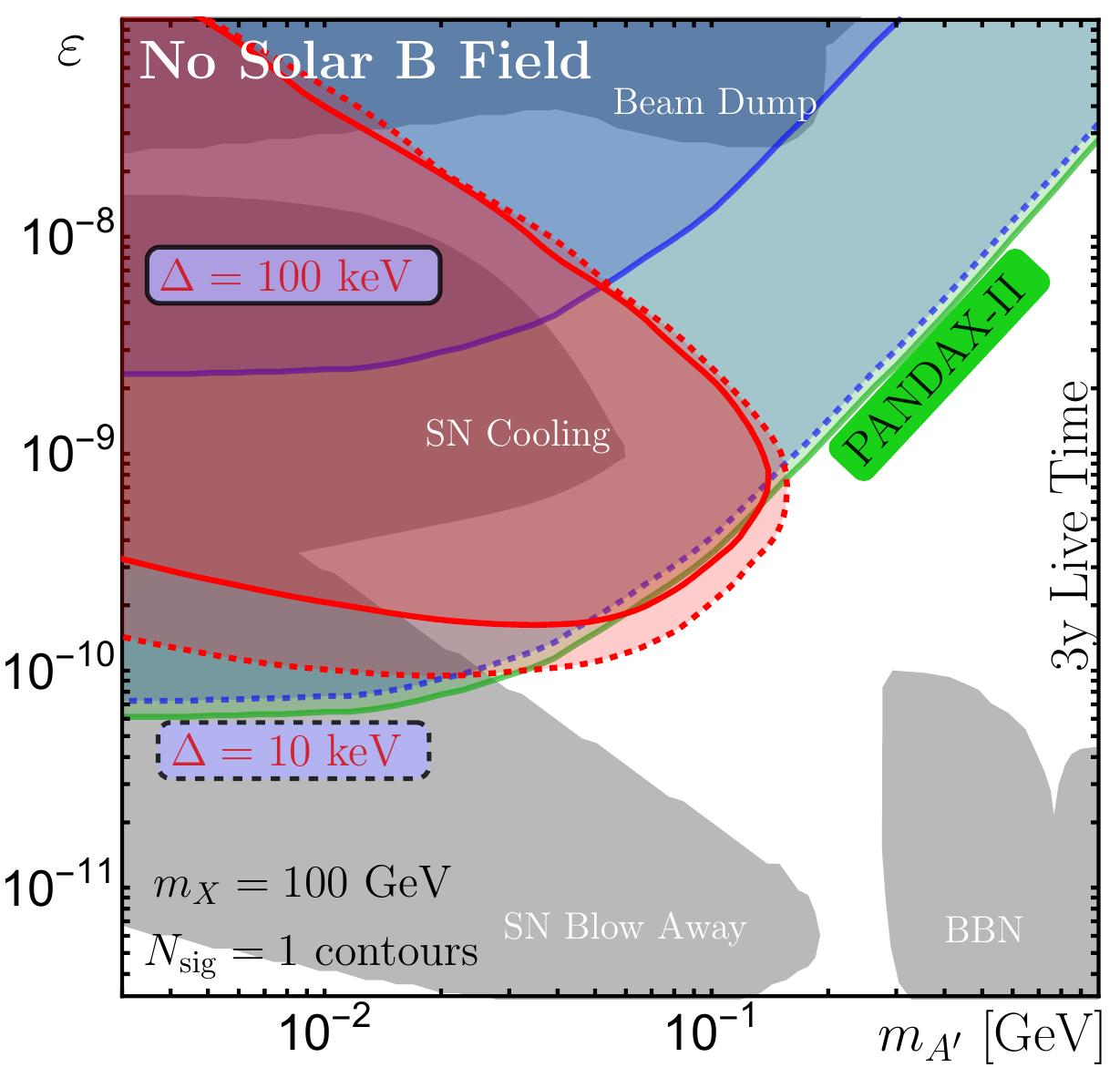} \qquad
	\includegraphics[width=0.45\linewidth]{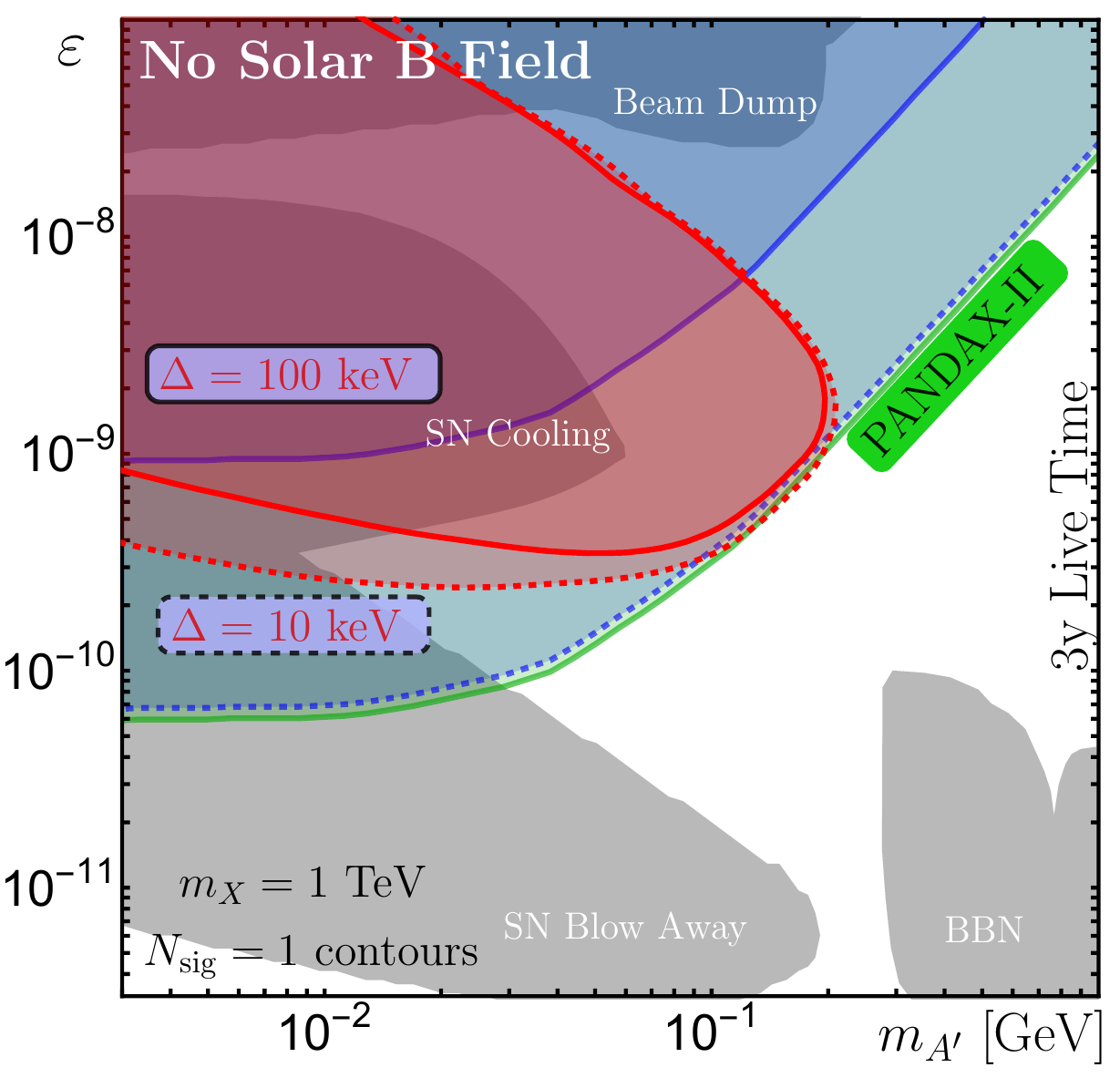} \\
	\includegraphics[width=0.45\linewidth]{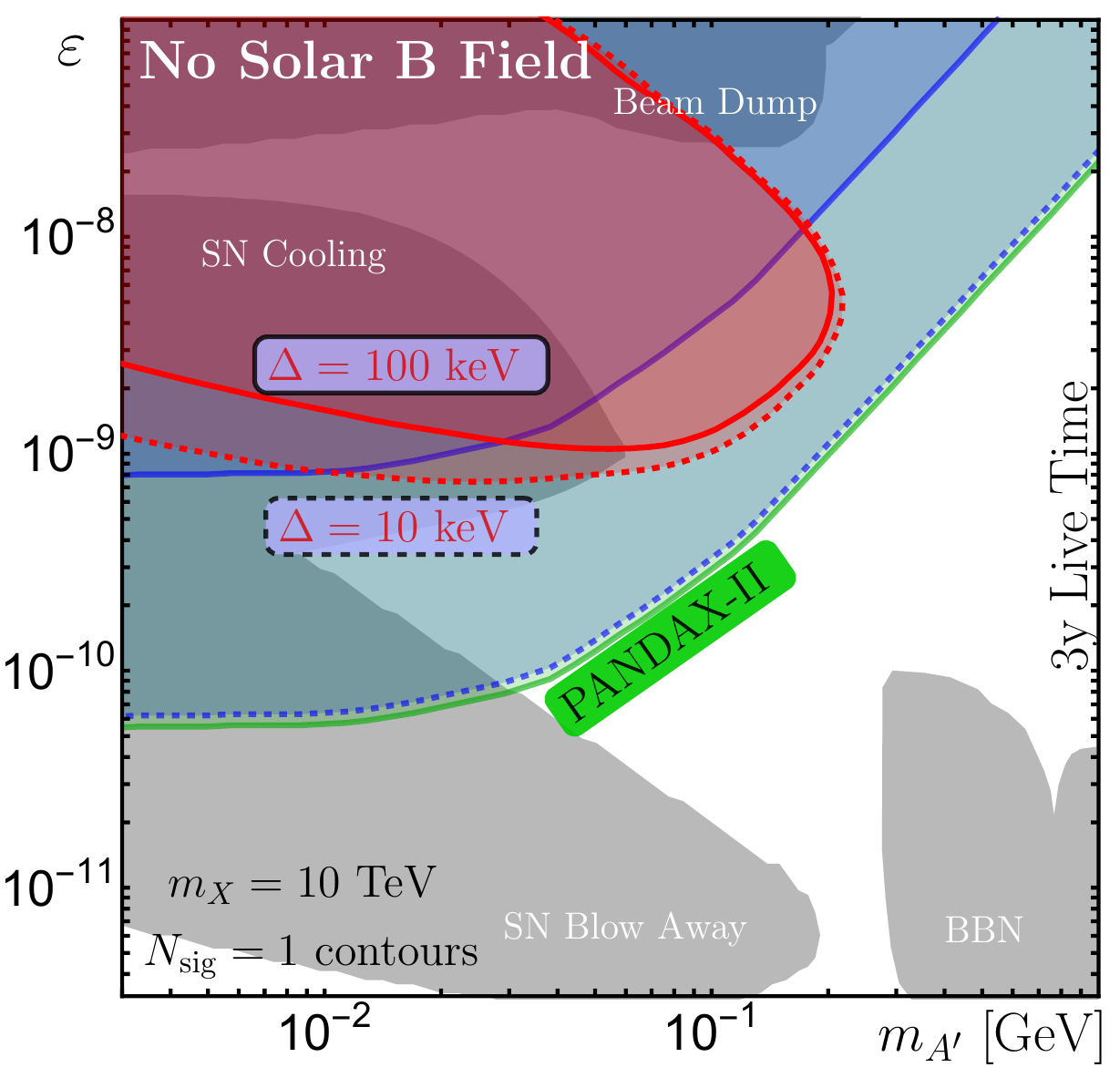} \qquad
	\includegraphics[width=0.45\linewidth]{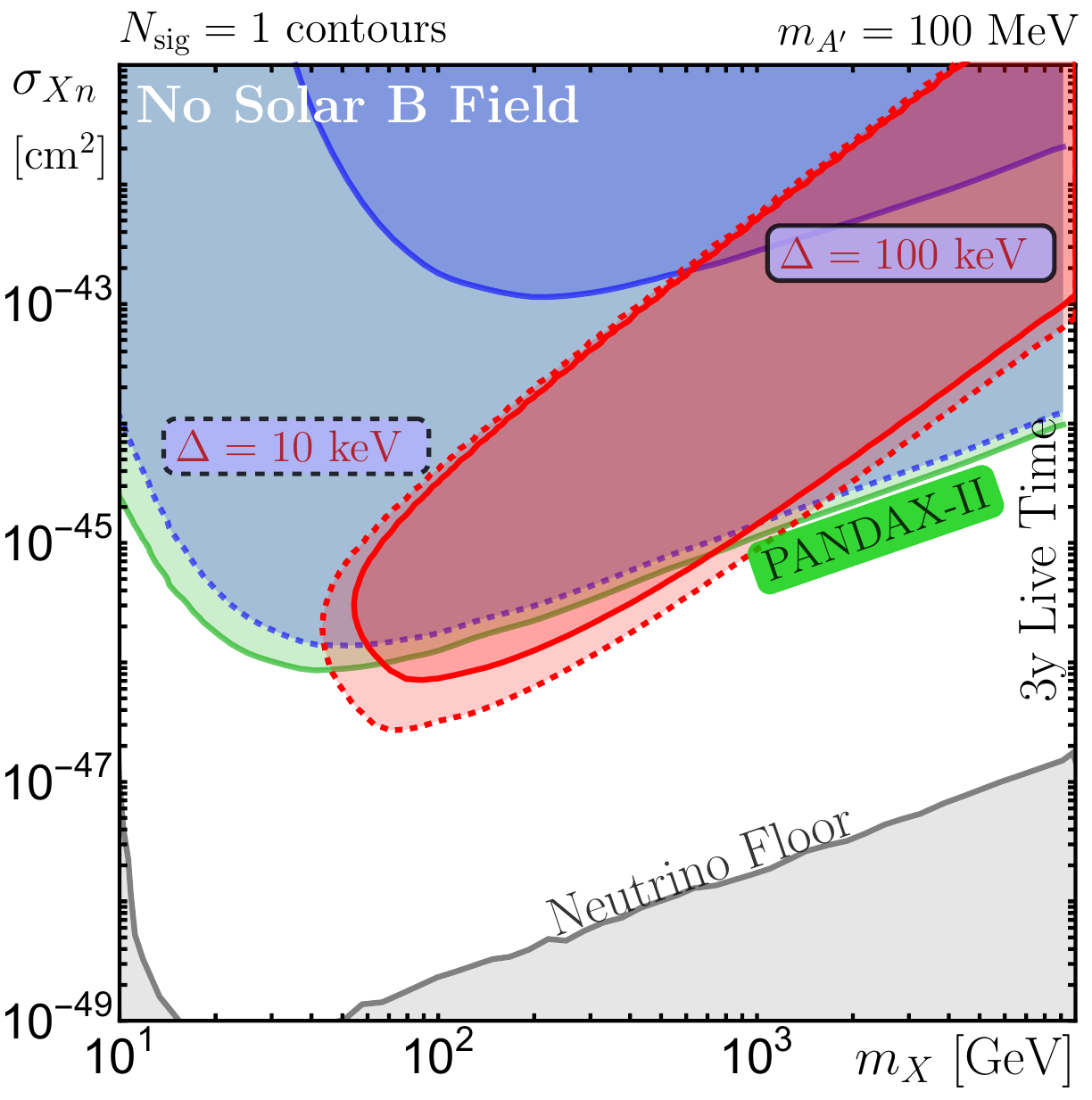}
	\vspace*{-0.1in} 
	\caption{Same as Figures \figsref{results}{resultshighexposure} but using the naive flux based estimate where magnetic deflections are ignored. }
	\label{fig:resultsnoB}
	\vspace*{-0.1in}
\end{figure}

\section{Results and Discussion}
\label{sec:results}

We present here the results of our analysis for the benchmark masses $m_X = 100 ~\gev$, $1 ~\tev$, and $10 ~\tev$ with mass splittings $\Delta = 10 ~\kev$ and $100 ~\kev$. When the mass splitting is lower than $10 ~\kev$ we find that the region that can be probed by AMS-02 is qualitatively indistinguishable from that of previous analyses, where the mass splitting was set to zero. An analysis of the current AMS-02 dataset will serve as a complementary probe to the most recent exclusions from the LUX direct detection experiment for dark sectors in which the mass splitting is $\Delta \lesssim 10~\kev$. For larger mass splittings, the AMS-02 signal region shrinks only modestly while exclusions from direct detection are relaxed by an order of magnitude or more in $\varepsilon$, leaving a significant window of parameter space that might only ever be probed by this search. The reduction of the AMS-02 signal window is reflective of the mild suppression of the capture rate: these contours do not approach the region of parameter space where the Sun is not presently in equilibrium and therefore the annihilation rate is directly proportional to the capture rate.

The reach of AMS-02 compared to other experimental probes discussed in \secref{exp. constraints}, is shown in \figref{results}. For clarity, we have shown only the $N_\text{sig} = 1$ contours for each value of the mass splitting. Contours corresponding to higher values of $N_\text{sig}$ are determined by the variation of $P_\text{det}$ as a function of $\varepsilon$ and $m_{A'}$, as discussed in \cite{Feng:2016ijc}. A rule of thumb for this, visible in our previous analysis, is that starting from the bottom of the $N_\text{sig} = 1$ contours, an increase of half a decade in $\varepsilon$ corresponds to a two decade increase in $N_\text{sig}$. Accordingly, there are regions outside of the modified direct detection exclusions where the expected $N_\text{sig}$ at AMS-02 may be of order 100 over the three year livetime. These are not significance contours. A more detailed analysis will be necessary to determine exactly the region of parameter space excluded by AMS, if indeed no dark sunshine signature is detected. However, these contours still provide useful information for weak scale secluded dark matter searches: the positron background at AMS-02 drops quickly at higher energies, to the extent that the expected number of background positrons of energy greater than $1 ~\tev$ is negligible. A single energetic positron observed from the direction of the Sun can therefore be very significant. For $m_X = 10 ~\tev$ then, the $N_\text{sig} = 1$ contours characterize AMS-02's reach region, while higher event numbers are required to claim discovery at lower energies with accordingly higher backgrounds.

More charitable assumptions yield accordingly more optimistic results: following our analyses in \cite{Feng:2016ijc}, we present the expected reach region for a hypothetical experiment with a higher exposure than AMS-02 in \figref{resultshighexposure}. If a detector with the same technical specifications and livetime as AMS-02 were placed near the Earth in such a fashion as to always face the Sun, perhaps at Lagrange Point 1, its exposure $\xi_\odot^\text{high}$ would be greater than the current exposure of AMS-02 by a factor of $80$ \cite{Machate}. In the formulae presented in \secref{detection} this is accomplished by setting $T = 240 ~\text{yr}$.

We also present the AMS-02 reach regions in the case where the magnetic field deflections are ignored in \figref{resultsnoB}. Such a signal region may be viable with an improved understanding and mapping of the interplanetary magnetic field. Note that while the search regions for $m_X = 100 ~\gev$ and $1 ~\tev$ grow substantially as compared to the regions presented in \figref{results}, the $m_X = 10 ~\tev$ region does not perceptibly change. This is easy to understand: as the dark matter mass increases, the dark photons produced in dark matter annihilation are accordingly more energetic. These highly boosted dark photons are more likely to decay to produce more energetic positrons, which experience less deflection in the solar magnetic field so that $P_\text{det}^B$ is negligible.

\section{Conclusions}
In previous work, we presented a novel method to discover a self-interacting elastic dark sector whose dark photon kinetically mixes with the SM photon. Dark matter is captured by the Sun and annihilates to dark photons, which furnish an indirect detection signature when they decay to $e^+ e^-$ pairs. For dark matter masses above $1 ~\tev$ this signature benefits from reduced astrophysical background. In spite of the low background, the signal region from previous analysis was largely excluded by independent bounds from dark matter searches at PANDAX-II. 

In this manuscript we have extended our earlier work to examine a dark sector consisting of two nearly-degenerate states coupling inelastically to a dark photon that mediates interactions with the Standard Model. Such a model benefits from weakened direct detection constraints and is well-motivated by small scale structure observations.
Relative to the previously considered elastic case, we found that the inelastic dark matter capture rate is only mildly suppressed and the non-perturbative Sommerfeld enhancement agrees with the elastic case, while the detection efficiencies are completely unchanged. As such, the region of parameter space over which the Sun is in equilibrium shrinks, but the region accessible to a dark sunshine search is largely determined by the detection efficiencies and thus is nearly unaffected. In contrast, direct detection bounds from LUX are relaxed by about an order of magnitude in $\varepsilon$. This leaves a region of parameter space: $1 ~\tev \lesssim m_X \lesssim 10 ~\tev$, $\Delta \sim 100 ~\kev$, $10 ~\mev \lesssim m_{A'} \lesssim 100 ~\mev$, and $10^{-10} \lesssim \varepsilon \lesssim 10^{-8}$, that is unprobed by supernova observations and fixed target dark photon searches, and favored to resolve small scale structure problems \cite{Blennow:2016gde}, where an inelastic dark sector may still be discovered or excluded using existing experiments and data. Finally, we provided estimates of the parameter space accessible to potential future experiments.

\section*{Acknowledgments} 
We are grateful to Fabian Machate and
Stefan Schael for providing us with their results on the exposure of AMS to the Sun. 
We
thank 
Adam Leibovich, Cameron Mahoney, and Andrew Zentner for discussions of supernova bounds on dark photons.
We also thank Jonathan Feng for helpful discussions and review of this manuscript.
We thank Adam Green for pointing out a typo in our decay length code, which moved the region of experimental sensitivity to values of $\varepsilon$ that are lower by an order of magnitude.
J.S.~thanks TASI 2016 ``Anticipating the Next Discoveries in Particle Physics'' at the University of Colorado, Boulder for its hospitality.
This work is supported by NSF Grant No.~PHY--1620638.  
J.S.~is supported by the Department of Education GAANN grant number P200A150121 at UCI.
Numerical calculations were
performed using \emph{Mathematica 10.4}~\cite{Mathematica10}.

\appendix*

\bibliography{bibinelastic}

\end{document}